\RequirePackage{fix-cm}
\documentclass[12pt]{article}
\usepackage[latin9]{inputenc}
\usepackage{geometry}
\usepackage{array}
\usepackage{float}
\usepackage{booktabs}
\usepackage{mathrsfs}
\usepackage{mathtools}
\usepackage{bm}
\usepackage{multirow}
\usepackage{algorithm2e}
\usepackage{amsmath}
\usepackage{amsthm}
\usepackage{amssymb}
\usepackage{graphicx}
\usepackage{setspace}
\usepackage[authoryear]{natbib}
\usepackage[unicode=true,
 bookmarks=false,
 breaklinks=true,pdfborder={0 0 1},backref=section,colorlinks=false]
 {hyperref}
\hypersetup{
 linkcolor=black}

\makeatletter

\providecommand{\tabularnewline}{\\}
\newcommand{\lyxdot}{.}

\usepackage{verbatim}
\usepackage{array}
\usepackage{tabularx}
\usepackage{ragged2e}
\newcolumntype{P}[1]{>{\RaggedRight\arraybackslash}p{#1}}

\usepackage[labelfont={bf}, 
            textfont={it},
            margin=1cm,
            font=small]{caption}
\usepackage[final]{microtype}
\usepackage{times}

\usepackage{titlesec}
\titleformat{\section}[block]{\bfseries\filcenter}{\thesection.}{1em}{}
\titleformat{\subsection}[hang]{\bfseries}{\thesubsection.}{1em}{}

\usepackage[multiple]{footmisc}

\LinesNumbered
\RestyleAlgo{algoruled} 

\global\long\def\ddd{,\ldots,}

\makeatother

\begin{document}

\title{\textbf{Tweedie Gradient Boosting for Extremely Unbalanced Zero-inflated
Data}}

\author{He Zhou\thanks{School of Statistics, University of Minnesota (zhou1354@umn.edu)},
Wei Qian\thanks{Department of Applied Economics and Statistics, University of Delaware
(weiqian@udel.edu)} and Yi Yang\thanks{Corresponding author, Department of Mathematics and Statistics, McGill
University (yi.yang6@mcgill.ca)} }
\maketitle
\begin{abstract}
Tweedie's compound Poisson model is a popular method to model insurance
claims with probability mass at zero and nonnegative, highly right-skewed
distribution. In particular, it is not uncommon to have extremely
unbalanced data with excessively large proportion of zero claims,
and even traditional Tweedie model may not be satisfactory for fitting
the data. In this paper, we propose a boosting-assisted zero-inflated
Tweedie model, called EMTboost, that allows zero probability mass
to exceed a traditional model. We makes a nonparametric assumption
on its Tweedie model component, that unlike a linear model, is able
to capture nonlinearities, discontinuities, and complex higher order
interactions among predictors. A specialized Expectation-Maximization
algorithm is developed that integrates a blockwise coordinate descent
strategy and a gradient tree-boosting algorithm to estimate key model
parameters. We use extensive simulation and data analysis on synthetic
zero-inflated auto-insurance claim data to illustrate our method's
prediction performance. 

\end{abstract}

\section{INTRODUCTION}

Setting premium for policyholders is one of the most important problems
in insurance business, and it is crucial to predict the size of actual
but unforeseeable claims. For typical portfolios in property and casualty
insurance business, the policy claim for a covered risk usually has
a highly right-skewed continuous distribution for positive claims,
while having a probability mass at zero when a claim does not occur.
This phenomenon poses unique challenges for data analysis as the data
cannot be transformed to normality by power transformation and special
treatment on zero claims is often required. In particular, \citet{jorgensen1994fitting}
and \citet{smyth2002fitting} used generalized linear models (GLM;
\citealp{nelder1972generalized}) with a Tweedie distributed outcome,
assuming Possion arrival of claims and Gamma distributed amount for
individual claims, to simultanuously model frequency and severity
of insurance claims. Although Tweedie GLM has been widely used in
actuarial studies (e.g., \citealp{mildenhall1999systematic}; \citealp{murphy2000using};
\citealp{sandri2008bias}), its structure of the logarithmic mean
is restricted to a linear form, which can be too rigid for some applications.
\citet{zhang2011cplm} modeled the nonlinearity by adding splines
to capture nonlinearity in claim data, and generalized additive models
(GAM; \citealp{hastie1990generalized}; \citealp{wood2006generalized})
can also model nonlinearity by estimating smooth functions. The structures
of these models have to be determined \textit{a priori} by specifying
spline degrees, main effects and interactions to be used in the model
fitting. More flexibly, \citet{yang2017insurance} proposed a nonparametric
Tweedie model to identify important predictors and their interactions. 

Despite the popularity of the Tweedie model under linear or nonlinear
logarithmic mean assumptions, it remains under-studied for problems
of modeling extremely unbalanced (zero-inflated) claim data. However,
it is well-known that the percentage of zeros in insurance claim data
can often be well over 90\%, posing challenges even for traditional
Tweedie model. In statistics literature, there are two general approaches
to handle data sets with excess zeros: the ``Hurdle-at-zero'' models
and the ``zero-inflated'' models. The Hurdle models (e.g., \citealp{cragg1971some};
\citealp{mullahy1986specification}) use a truncated-at-zero strategy,
whose examples include truncated Poisson and truncated negative-binomial
models. On the other hand, ``zero-inflated'' models typically use
a mixture model strategy, whose examples include zero-inflated Poisson
regression and zero-inflated negative binomial regression (e.g., \citealp{lambert1992zero};
\citealp{hall2000zero}; \citealp{frees2016multivariate}), among
many notable others. 

In this paper, we aim to tackle the entremely unbalanced insurance
data problem with excessive zeros by developing a zero-inflated nonparametric
Tweedie compound Poisson model. To our knowledge, no existing work
systematically studied the zero-inflated Tweedie model and its computational
issues. Under a mixture model framework that subsumes traditional
Tweedie model as a special case, we develop an Expectation-Maximization
(EM) algorithm that efficiently integrates a blockwise coordinate
descent algorithm and a gradient boosting-type algorithm to estimate
key parameters. We call our method as EMTboost for brevity.

The EMTboost method assumes a mixture of Tweedie model component and
a mass zero component. As one interesting feature, it can simultaneously
provide estimation for the zero mass probability as well as the dispersion/power
parameters of the Tweedie model component, which are useful information
in understanding the zero-inflated nature of claim data under analysis.
In addition, we employ boosting techniques to fit the mean of the
Tweedie component. This boosting approach is motivated by its proven
success for nonparametric regression and classification (\citealp{freund1997decision};
\citealp{breiman1998arcing}; \citealp{breiman1999prediction}; \citealp{friedman2001greedy},\citeyear{friedman2002stochastic},\citeyear{friedman2001elements}).
By integrating a gradient-boosting algorithm with trees as weak learners,
the zero-inflated model can learn nonlinearities, discontinuities
and complex higher order interactions of predictors, and potentially
reduce modeling bias to produce high predictive performance. Due to
the inherent use trees, this approach also naturally handles missing
values, outliers and various predictor types. 

The rest of the article is organized as follows. Section \ref{section:TD and ZIF}
briefly present the models. The main methodology with implementation
details is given in Section \ref{section:EMTboost}. We use simulation
to show performance of EMTboost in Section \ref{section:Simulation Studies},
and apply it to analyze an auto-insurance claim data in Section \ref{section:application:real data}.
Brief concluding remarks are given in Section \ref{section:Conclusions}.

\section{ZERO-INFLATED TWEEDIE MODEL\label{section:TD and ZIF}}

To begin with, we give a brief overview of the Tweedie's compound
Poisson model, followed by the introduction of the zero-inflated Tweedie
model. Let $N$ be a Poisson random variable denoted by $\mathrm{Pois}(\lambda)$
with mean $\lambda$, and let $\tilde{Z}_{d}$'s ($d=0,1,\dots,N$)
be i.i.d Gamma random variables denoted by $\mathrm{Gamma}(\alpha,\gamma)$
with mean $\alpha\gamma$ and variance $\alpha\gamma^{2}$. Assume
$N$ is independent of $\tilde{Z}_{d}$'s. Define a compound Poisson
random variable $Z$ by
\begin{equation}
Z=\left\{ \begin{array}{ll}
0 & \text{if }N=0,\\
\tilde{Z}_{1}+\tilde{Z}_{2}+\cdots+\tilde{Z}_{N} & \text{if }N=1,2,\dots
\end{array}\right..
\end{equation}
Then $Z$ is a Poisson sum of independent Gamma random variables.
The compound Poisson distribution (\citealp{jorgensen1994fitting};
\citealp{smyth2002fitting}) is closely connected to a special class
of exponential dispersion models (\citealp{jorgensen1987exponential})
known as Tweedie models \citep{tweedie1984index}, whose probability
density functions are of the form
\begin{equation}
f_{Z}(z|\theta,\phi)=a(z,\phi)\exp\left\{ \frac{z\theta-\kappa(\theta)}{\phi}\right\} ,\label{EDM}
\end{equation}
where $a(\cdot)$ and $\kappa(\cdot)$ are given functions, with $\theta\in\mathbb{R}$
and $\phi\in\mathbb{R}^{+}$. For Tweedie models, the mean and variance
of $Z$ has the property $\mathrm{\mathbb{E}}(Z)\coloneqq\mu=\dot{\kappa}(\theta),~\mathrm{Var}(Z)=\phi\ddot{\kappa}(\theta)$,
where $\dot{\kappa}(\theta)$ and $\ddot{\kappa}(\theta)$ are the
first and second derivatives of $\kappa(\theta)$, respectively. The
power mean-variance relationship is $\mathrm{Var}(Z)=\phi\mu^{\rho}$
for some index parameter $\rho\in(1,2)$, which gives $\theta=\mu^{1-\rho}/(1-\rho)$,
$\kappa(\theta)=\mu^{2-\rho}/(2-\rho)$ and $\ddot{\kappa}(\theta)=\mu^{\rho}$.
If we re-parameterize the compound Poisson model by
\begin{equation}
\lambda=\frac{1}{\phi}\frac{\mu^{2-\rho}}{2-\rho},~~\alpha=\frac{2-\rho}{\rho-1},~~\gamma=\phi(\rho-1)\mu^{\rho-1},
\end{equation}
then it will have the form of a Tweedie model $\textrm{Tw}(\mu,\phi,\rho)$
with the probability density function
\begin{equation}
f_{\text{Tw}}(z|\mu,\phi,\rho)\coloneqq a(z,\phi,\rho)\exp\left(\frac{1}{\phi}\left(z\frac{\mu^{1-\rho}}{1-\rho}-\frac{\mu^{2-\rho}}{2-\rho}\right)\right)\label{Tweedie pdf}
\end{equation}
where
\begin{equation}
a(z,\phi,\rho)=\left\{ \begin{array}{ll}
1, & \text{if }z=0,\\
\frac{1}{z}\sum_{t=1}^{\infty}W_{t}(z,\phi,\rho)\\
~~=\frac{1}{z}\sum_{t=1}^{\infty}\frac{z^{t\alpha}}{(\rho-1)^{t\alpha}(2-\rho)^{t}\Gamma(t\alpha)\phi^{t(1+\alpha)}t!}, & \text{if }z>0,
\end{array}\right.\label{infinite sum function}
\end{equation}
with $\alpha=(2-\rho)/(\rho-1)$. When $z>0$, the sum of infinite
series $\sum_{t=1}^{\infty}W_{t}$ is an example of Weight's generalized
Bessel function.

With the formulation above, the Tweedie model has positive probability
mass at zero with $\mathrm{\mathbb{P}}(Z=0)=\mathrm{\mathbb{P}}(N=0)=\exp(-\lambda)$.
Despite its popularity in actuarial studies, Tweedie models do not
always give ideal performance in cases when the empirical distribution
of claim data (e.g., in auto insurance), is extremely unbalanced and
has an excessively high proportion of zero claims, which will be illustrated
in the numerical exposition. This motivates us to consider a zero-inflated
mixture model that combines a Tweedie distribution with probability
$q$ and an exact zero mass with probability $1-q$ :
\begin{equation}
Y=\left\{ \begin{array}{ll}
Z, & \text{with probability }q,\ \text{where\ }Z\sim\text{Tw}(\mu,\phi,\rho),\\
0, & \text{with probability }1-q.
\end{array}\right.\label{ZIF model}
\end{equation}
We denote this zero-inflated Tweedie model by $Y\sim\text{ZIF-Tw}(\mu,\phi,\rho,q)$.
The probability density function of $Y$ can be written as
\begin{equation}
f_{\text{ZIF-Tw}}(y|\mu,\phi,\rho,q)\coloneqq qf_{\text{Tw}}(y|\mu,\phi,\rho)+(1-q)I\left\{ y=0\right\} ,
\end{equation}
so that $\mathbb{P}\left(Y=0\right)=q\exp\left(-\frac{1}{\phi}\frac{\mu^{2-\rho}}{2-\rho}\right)+\left(1-q\right)$
and $\mathbb{E}\left(Y\right)=q\mu$.

\section{METHODOLOGY\label{section:EMTboost}}

Consider a portfolio of polices $\mathbf{D}=\{(y_{i},\bm{\mathrm{x}}_{i},\omega_{i})\}_{i=1}^{n}$
from $n$ independent insurance policy contracts, where for the $i$-th
contract, $y_{i}$ is the policy pure premium, $\bm{\mathrm{x}}_{i}$
is a $p$-dimensional vector of explanatory variables that characterize
the policyholder and the risk being insured, and $\omega_{i}$ is
the policy duration, i.e., the length of time that the policy remains
in force. Assume that each policy pure premium $y_{i}$ is an observation
from the zero-inflated Tweedie distribution $Y_{i}\sim\text{ZIF-Tw}(\mu_{i},\phi/\omega_{i},\rho,q)$
as defined in \eqref{ZIF model}. For now we assume that the value
of $\rho$ is given and in the end of this section we will discuss
the estimation of $\rho$. Assume $\mu_{i}$ is determined by a regression
function $F:\mathbb{R}^{p}\rightarrow\mathbb{R}$ of $\bm{\mathrm{x}}_{i}$
through the $\log$ link function
\begin{equation}
\log(\mu_{i})=\log\left\{ \mathrm{\mathbb{E}}\left(Y_{i}|\bm{\mathrm{x}}_{i}\right)\right\} =F(\bm{\mathrm{x}}_{i}).
\end{equation}
Let $\boldsymbol{\theta}=\left(F,\phi,q\right)\in\mathcal{F}\times\mathbb{R}^{+}\times\left[0,1\right]$
denote a collection of parameters to be estimated with $\mathcal{F}$
denoting a class of regression functions (based on tree learners).
Our goal is to maximize the log-likelihood function of the mixture
model
\begin{equation}
\widehat{\boldsymbol{\theta}}=\arg\max_{\boldsymbol{\theta}\in\mathbf{\Theta}}\log\mathcal{L}\left(\boldsymbol{\theta};\mathbf{D}\right).\label{eq:likelihood}
\end{equation}
where
\begin{align}
\mathcal{L}\left(\boldsymbol{\theta};\mathbf{D}\right)\coloneqq\prod_{i=1}^{n}f_{\text{ZIF-Tw}}\left(y_{i}|\exp\left(F(\bm{\mathrm{x}}_{i})\right),\phi/\omega_{i},\rho,q\right) & ,\label{ZIFlikelihood}
\end{align}
but doing so directly is computationally difficult. To efficiently
estimate $\boldsymbol{\theta}=\left(F,\phi,q\right)$, we propose
a gradient-boosting based EM algorithm, referred to as EMTboost henceforth.
We first give an outline of the EMTboost algorithm and the details
will be discussed further in Section \ref{subsection:TDboost}--\ref{subsection:modifed EM}.
The basic idea is to first construct a proxy $Q\text{-function}$
corresponding to the current iterate by which the target likelihood
function \eqref{ZIFlikelihood} is lower bounded (E-step), and then
maximize the $Q\text{-function}$ to get the next update (M-step)
so that \ref{ZIFlikelihood} can be driven uphill:

\paragraph*{E-Step Construction }

\noindent We introduce $n$ independent Bernoulli latent variables
$\Pi_{1},\ldots,\Pi_{n}$ such that for $i=1,\ldots,n$, $P(\Pi_{i}=1)=q$,
and $\Pi_{i}=1$ when $y_{i}$ is sampled from $\text{Tweedie}(\mu_{i},\phi/\omega_{i},\rho)$
and $\Pi_{i}=0$ if $y_{i}$ is from the exact zero point mass. Denote
$\boldsymbol{\Pi}=(\Pi_{1},\ldots,\Pi_{n})^{\top}$. Assume predictors
$\mathbf{x}_{i}$'s are fixed. Given $\Pi_{i}\in\left\{ 0,1\right\} $
and $\boldsymbol{\theta}$, the joint-distribution of the complete
model for each observation is
\begin{equation}
f\left(y_{i},\Pi_{i}|\boldsymbol{\theta}\right)\coloneqq\left(q\cdot f_{\text{Tw}}\left(y_{i}|\exp\left(F(\bm{\mathrm{x}}_{i})\right),\phi,\omega_{i}\right)\right)^{\Pi_{i}}\left(\left(1-q\right)\cdot I\left\{ y_{i}=0\right\} \right)^{1-\Pi_{i}}.\label{joint-distribution}
\end{equation}
The posterior distribution of each latent variable $\Pi_{i}$ is 
\begin{equation}
f\left(\Pi_{i}|y_{i},\boldsymbol{\theta}\right)=\frac{f\left(y_{i},\Pi_{i}|\boldsymbol{\theta}\right)}{f\left(y_{i},\Pi_{i}|\boldsymbol{\theta}\right)+f\left(y_{i},1-\Pi_{i}|\boldsymbol{\theta}\right)}.\label{posterior distribution}
\end{equation}
For the E-step construction, denote $\boldsymbol{\theta}^{t}=\left(F^{t},\phi^{t},q^{t}\right)$
the value of $\boldsymbol{\theta}$ during $t$-th iteration of the
EMTboost algorithm. The $Q\text{-function}$ for each observation
is
\begin{align}
Q_{i}\left(\boldsymbol{\theta}|\boldsymbol{\theta}^{t}\right) & \coloneqq\mathbb{E}_{\Pi_{i}\sim f\left(\Pi_{i}|y_{i},\boldsymbol{\theta}^{t}\right)}\left[\log f\left(y_{i},\Pi_{i}|\boldsymbol{\theta}\right)\right]\nonumber \\
 & =f\left(\Pi_{i}=1|y_{i},\boldsymbol{\theta}^{t}\right)\log f\left(y_{i},\Pi_{i}=1|\boldsymbol{\theta}\right)+f\left(\Pi_{i}=0|y_{i},\boldsymbol{\theta}^{t}\right)\log f\left(y_{i},\Pi_{i}=0|\boldsymbol{\theta}\right)\nonumber \\
 & =\delta_{1,i}^{t}(\boldsymbol{\theta}^{t})\log\left(q\cdot f_{\text{Tw}}\left(y_{i}|\exp\left(F(\textbf{x}_{i})\right),\phi,\omega_{i}\right)\right)+\delta_{0,i}^{t}(\boldsymbol{\theta}^{t})\log\left(1-q\right)I\left\{ y_{i}=0\right\} ,\label{point-wise Q-function}
\end{align}
where
\begin{align}
\delta_{1,i}^{t} & (\boldsymbol{\theta}^{t})=f\left(\Pi_{i}=1|y_{i},\boldsymbol{\theta}^{t}\right)=\left\{ \begin{array}{ll}
1, & \text{if }y_{i}>0;\\
\frac{q^{t}\exp\left(\frac{\omega_{i}}{\phi^{t}}\left(-\frac{\exp\left(F^{t}\left(\textbf{x}_{i}\right)\left(2-\rho\right)\right)}{2-\rho}\right)\right)}{q^{t}\exp\left(\frac{\omega_{i}}{\phi^{t}}\left(-\frac{\exp\left(F^{t}\left(\textbf{x}_{i}\right)\left(2-\rho\right)\right)}{2-\rho}\right)\right)+\left(1-q^{t}\right)}, & \text{if }y_{i}=0,
\end{array}\right.\label{posterior density1}\\
\delta_{0,i}^{t} & (\boldsymbol{\theta}^{t})=f\left(\Pi_{i}=0|y_{i},\boldsymbol{\theta}^{t}\right)=1-\delta_{1,i}^{t}(\boldsymbol{\theta}^{t})\label{posterior density0}
\end{align}
Given $n$ observations of data $\mathbf{D}=\{(y_{i},\bm{\mathrm{x}}_{i},\omega_{i})\}_{i=1}^{n}$,
the $Q\text{-function}$ is{\small{}
\begin{align}
Q\left(\boldsymbol{\theta}|\boldsymbol{\theta}^{t}\right) & =\frac{1}{n}\sum_{i=1}^{n}Q_{i}\left(\boldsymbol{\theta}|\boldsymbol{\theta}^{t}\right)\nonumber \\
 & =\frac{1}{n}\sum_{i=1}^{n}\delta_{1,i}^{t}(\boldsymbol{\theta}^{t})\log\left(q\cdot f_{\text{Tw}}\left(y_{i}|\exp\left(F(\bm{\mathrm{x}}_{i})\right),\phi,\omega_{i}\right)\right)+\delta_{0,i}^{t}(\boldsymbol{\theta}^{t})\log\left(\left(1-q\right)\cdot I\left\{ y_{i}=0\right\} \right)\nonumber \\
 & =\frac{1}{n}\sum_{i=1}^{n}\delta_{1,i}^{t}(\boldsymbol{\theta}^{t})\log\left\{ qa(y_{i},\phi/\omega_{i},\rho)\exp\left[\frac{\omega_{i}}{\phi}\left(y_{i}\frac{\exp\left((1-\rho)F(\bm{\mathrm{x}}_{i})\right)}{1-\rho}-\frac{\exp\left((2-\rho)F(\bm{\mathrm{x}}_{i})\right)}{2-\rho}\right)\right]\right\} \nonumber \\
 & +\frac{1}{n}\sum_{\{i:y_{i}=0\}}\delta_{0,i}^{t}(\boldsymbol{\theta}^{t})\log\left(1-q\right)\label{Q-function}
\end{align}
}{\small\par}

\paragraph*{M-Step Maximization}

\noindent Given the $Q\text{-function}$ \eqref{Q-function}, we update
$\boldsymbol{\theta}^{t}$ to $\boldsymbol{\theta}^{t+1}$ through
maximization of \eqref{Q-function} by
\begin{equation}
\boldsymbol{\theta}^{t+1}=\left(F^{t+1},\phi^{t+1},q^{t+1}\right)\longleftarrow\arg\max_{\boldsymbol{\theta}\in\boldsymbol{\Theta}}Q\left(\boldsymbol{\theta}|\boldsymbol{\theta}^{t}\right),\label{M-step maximizer}
\end{equation}
in which $F^{t+1}$, $\phi^{t+1}$ and $q^{t+1}$ are updated successively
through blockwise coordinate descent 
\begin{align}
F^{t+1} & \longleftarrow\arg\max_{F\in\mathcal{F}}Q\left(F|\left(F^{t},\phi^{t},q^{t}\right)\right)\label{eq:line1}\\
\phi^{t+1} & \longleftarrow\arg\max_{\phi\in\mathbb{R}^{+}}Q\left(\phi|\left(F^{t+1},\phi^{t},q^{t}\right)\right)\label{eq:line2}\\
q^{t+1} & \longleftarrow\arg\max_{q}Q\left(q|\left(F^{t+1},\phi^{t+1},q^{t}\right)\right)\label{eq:line3}
\end{align}
Specifically, \eqref{eq:line1} is equivalent to update
\begin{align}
F^{t+1} & \longleftarrow\arg\max_{F\in\mathscr{F}}\sum_{i=1}^{n}\delta_{1,i}^{t}(F^{t},\phi^{t},q^{t})\Psi(y_{i},F(\bm{\mathrm{x}}_{i}),\omega_{i}),\label{update:F}
\end{align}
where the risk function $\Psi$ is defined as
\begin{equation}
\Psi(y_{i},F(\bm{\mathrm{x}}_{i}),\omega_{i})=\omega_{i}\left(y_{i}\frac{\exp\left(F\left(\bm{\mathrm{x}}_{i}\right)\left(1-\rho\right)\right)}{1-\rho}-\frac{\exp\left(F\left(\bm{\mathrm{x}}_{i}\right)\left(2-\rho\right)\right)}{2-\rho}\right).\label{i-th risk function}
\end{equation}
We use a gradient tree-boosted algorithm to compute \eqref{update:F},
and its details are deferred to Section \ref{subsection:TDboost}.
After updating $F^{t+1}$ we then update $\phi^{t+1}$ in \eqref{eq:line2}
using
\begin{align}
\phi^{t+1} & \longleftarrow\arg\max_{\phi\in\mathbb{R}^{+}}\sum_{i=1}^{n}\delta_{1,i}^{t}(F^{t+1},\phi^{t},q^{t})\Biggl\{\log a(y_{i},\phi/\omega_{i},\rho),\label{update:phi}\\
 & +\frac{\omega_{i}}{\phi}\left(y_{i}\frac{\exp\left((1-\rho)F^{t+1}(\bm{\mathrm{x}}_{i})\right)}{1-\rho}-\frac{\exp\left((2-\rho)F^{t+1}(\bm{\mathrm{x}}_{i})\right)}{2-\rho}\right)\Biggl\}.\nonumber 
\end{align}

\noindent Conditional on the updated $F^{t+1}$ and $q^{t}$, maximizing
the log-likelihood function with respect to $\phi$ in \eqref{update:phi}
is a univariate optimization problem that can be solved by using a
combination of golden section search and successive parabolic interpolation
\citep{brent2013algorithms}. 

\noindent After updating $F^{t+1}$ and $\phi^{t+1}$, we can use
a simple formula to update $q^{t+1}$ for \eqref{eq:line3}
\begin{equation}
q^{t+1}\longleftarrow\frac{1}{n}\sum_{i=1}^{n}\delta_{1,i}^{t}\left(F^{t+1},\phi^{t+1},q^{t}\right).\label{update:q}
\end{equation}

\noindent We repeat the above E-step and M-step iteratively until
convergence. In summary, the complete EMTboost algorithm is shown
in Algorithm \ref{algorithm:EMTboost}. 

\noindent 
\begin{algorithm}
\SetKwData{Left}{left}\SetKwData{This}{this}\SetKwData{Up}{up}
\SetKwFunction{Union}{Union}\SetKwFunction{FindCompress}{FindCompress}
\SetKwInOut{Input}{Input}\SetKwInOut{Output}{Output}

\Input{Dataset $\mathbf{D}=\{(y_{i},\bm{\mathrm{x}}_{i},\omega_{i})\}_{i=1}^{n}$ and the index parameter $\rho$.}
\Output{Estimates $\widehat{\boldsymbol{\theta}}=(\widehat{F},\widehat{\phi},\widehat{q})$.} 

Initialize $\boldsymbol{\theta}^{0}=(F^{0},\phi^{0},\rho^{0})$. Compute the  index set $\mathcal{I}=\{i:y_{i}=0\}$ and initialize $\{\delta_{0,i}^{0},\delta_{1,i}^{0}\}_{i\in\mathcal{I}}$ by setting $\delta_{1i}=1,\delta_{0i}=0$      for  $i\notin\mathcal{I}$.\

\For{$t=0,1,2\ddd T$}{
\textbf{E-step:} Update $\{\delta_{0,i}^{t},\delta_{1,i}^{t}\}_{i\in\mathcal{I}}$ by \eqref{posterior density1} and \eqref{posterior density0}.

\textbf{M-step:} Update $\boldsymbol{\theta}^{t+1}=\left(F^{t+1},\phi^{t+1},q^{t+1}\right)$ by using \eqref{update:F} that calls Algorithm \ref{algorithm:TDboost}, \eqref{update:phi} and \eqref{update:q}.
\begin{align*}
F^{t+1} & \longleftarrow\arg\max_{F\in\mathcal{F}}Q\left(F|\left(F^{t},\phi^{t},q^{t}\right)\right) \\
\phi^{t+1} & \longleftarrow\arg\max_{\phi\in\mathbb{R}^{+}}Q\left(\phi|\left(F^{t+1},\phi^{t},q^{t}\right)\right) \\
q^{t+1} & \longleftarrow\arg\max_{q}Q\left(q|\left(F^{t+1},\phi^{t+1},q^{t}\right)\right) 
\end{align*}

}
Return $\widehat{\boldsymbol{\theta}}=(\widehat{F},\widehat{\phi},\widehat{q})=\left(F^{T},\phi^{T},q^{T}\right)$.

\caption{EMTboost Algorithm\label{algorithm:EMTboost}}
\end{algorithm}

So far we only assume that the value of $\rho$ is known when estimating
$\boldsymbol{\theta}=\left(F,\phi,q\right)$. Next we give a profile
likelihood method to jointly estimate\textbf{ $(\boldsymbol{\theta},\rho)=\left(F,\phi,q,\rho\right)$}
when $\rho$ is unknown. Following \citet{dunn2005series},\textbf{
}we pick a sequence of $K$ equally-spaced candidate values $\{\rho_{1},\cdots,\rho_{K}\}$
on the interval $(1,2)$, and for each fixed $\rho_{k}$, $k=1,\ldots,K$,
we apply Algorithm \ref{algorithm:EMTboost} to maximize the log-likelihood
function \eqref{ZIFlikelihood} with respect to $\boldsymbol{\theta}_{\rho_{k}}=\left(F_{\rho_{k}},\phi_{\rho_{k}},q_{\rho_{k}}\right)$,
which gives the corresponding estimators $\widehat{\boldsymbol{\theta}}_{\rho_{k}}=\left(\widehat{F}_{\rho_{k}},\widehat{\phi}_{\rho_{k}},\widehat{q}_{\rho_{k}}\right)$
and the log-likelihood function $\mathcal{L}\left(\widehat{\boldsymbol{\theta}}_{\rho_{k}};\mathbf{D},\rho_{k}\right)$.
Then from the sequence $\{\rho_{1},\cdots,\rho_{K}\}$ we choose the
optimal $\widehat{\rho}$ as the maximizer of $\mathcal{L}$.
\begin{equation}
\widehat{\rho}=\arg\max_{\rho\in\{\rho_{1},\cdots,\rho_{K}\}}\left\{ \mathcal{L}\left(\widehat{\boldsymbol{\theta}}_{\rho};\mathbf{D},\rho\right)\right\} .
\end{equation}
We then obtain the corresponding estimator $\widehat{\boldsymbol{\theta}}_{\widehat{\rho}}=\left(\widehat{F}_{\widehat{\rho}},\widehat{\phi}_{\widehat{\rho}},\widehat{q}_{\widehat{\rho}}\right)$.

\begin{algorithm}
\SetKwData{Left}{left}\SetKwData{This}{this}\SetKwData{Up}{up}
\SetKwFunction{Union}{Union}\SetKwFunction{FindCompress}{FindCompress}
\SetKwInOut{Input}{Input}\SetKwInOut{Output}{Output}

\Input{Dataset $\mathbf{D}=\{(y_{i},\bm{\mathrm{x}}_{i},\omega_{i})\}_{i=1}^{n}$.}
\Output{Estimates $\widehat{\boldsymbol{\theta}}_{\widehat{\rho}}=\left(\widehat{F}_{\widehat{\rho}},\widehat{\phi}_{\widehat{\rho}},\widehat{q}_{\widehat{\rho}}\right)$.} 
Pick a sequence of $K$ equally-spaced candidate values $\{\rho_{1},\cdots,\rho_{K}\}$ on the interval $(1,2)$. 
\For{$k=1\ddd K$}{
Set $\rho=\rho_k$.

Call Algorithm \ref{algorithm:EMTboost} to compute $\widehat{\boldsymbol{\theta}}_{\rho_{k}}=\left(\widehat{F}_{\rho_{k}},\widehat{\phi}_{\rho_{k}},\widehat{q}_{\rho_{k}}\right)$ and the corresponding log-likelihood function
$\mathcal{L}\left(\widehat{\boldsymbol{\theta}}_{\rho_{k}};\mathbf{D},\rho_{k}\right)$.
}
Compute the optimal $\widehat{\rho}$
$$\widehat{\rho}={\arg\max}_{\rho\in\{\rho_{1},\cdots,\rho_{K}\}}\left\{ \mathcal{L}\left(\widehat{\boldsymbol{\theta}}_{\rho};\mathbf{D},\rho\right)\right\}.$$

Return the final estimator $\widehat{\boldsymbol{\theta}}_{\widehat{\rho}}=\left(\widehat{F}_{\widehat{\rho}},\widehat{\phi}_{\widehat{\rho}},\widehat{q}_{\widehat{\rho}}\right)$.  

\caption{Profile Likelihood for EMTboost\label{algorithm:profile_likelihood}}
\end{algorithm}

\subsection{Estimating $F\left(\cdot\right)$ via Tree-based Gradient Boosting
\label{subsection:TDboost}}

To minimize the weighted sum of the risk function \eqref{i-th risk function},
we employ the tree-based gradient boosting algorithm to recover the
predictor function $F\left(\cdot\right)$ :
\begin{equation}
\widetilde{F}(\cdot)=\arg\min_{F(\cdot)\in\mathcal{F}}\sum_{i=1}^{n}\delta_{1,i}\Psi(y_{i},F(\bm{\mathrm{x}}_{i}),\omega_{i}),\label{empirical risk function}
\end{equation}
Note that the objective function does not depend on $\phi$. To solve
the gradient-tree boosting, each candidate function $F\in\mathcal{F}$
is assumed to be an ensemble of $L$-terminal nodes regression trees,
as base learners:
\begin{align}
F(\mathbf{x}) & =F^{[0]}+\sum_{m=1}^{M}\beta^{[m]}h\left(\bm{\mathrm{x}};\bm{\xi}^{[m]}\right),\label{ensemble of base learners}\\
 & =F^{[0]}+\sum_{m=1}^{M}\beta^{[m]}\left\{ \sum_{l=1}^{L}u_{l}^{[m]}I(\bm{\mathrm{x}}\in R_{l}^{[m]})\right\} 
\end{align}
where $F^{[0]}$ is a constant scalar, $\beta^{[m]}$ is the expansion
coefficient and $h(\bm{\mathrm{x}};\bm{\xi}^{[m]})$ is the $m$-th
base learner, characterized by the parameter $\bm{\xi}^{[m]}=\{R_{l}^{[m]},u_{l}^{[m]}\}_{l=1}^{L}$
, with $R_{l}^{[m]}$ being the disjoint regions representing the
terminal nodes of the tree, and constants $u_{l}^{[m]}$ being the
values assigned to each region. 

The constant $\widehat{F}^{[0]}$ is chosen as the $1$-terminal tree
that minimizes the negative log-likelihood. A forward stagewise algorithm
\citep{friedman2001greedy} builds up the components $\beta^{[m]}h\left(\bm{\mathrm{x}};\bm{\xi}^{[m]}\right)$
sequentially through a gradient-descent-like approach with $m=1,2,\dots,M$.
At iteration stage $m$, suppose that the current estimation for $\widetilde{F}\left(\cdot\right)$
is $\widehat{F}^{[m-1]}\left(\cdot\right)$. To update from $\widehat{F}^{[m-1]}\left(\cdot\right)$
to $\widehat{F}^{[m]}\left(\cdot\right)$, the gradient-tree boosting
method fits the $m$-th regression tree $h\left(\bm{\mathrm{x}};\bm{\xi}^{[m]}\right)$
to the negative gradient vector by least-squares function minimization:
\begin{equation}
\widehat{\bm{\xi}}^{\left[m\right]}=\arg\min_{\boldsymbol{\xi}^{\left[m\right]}}\sum_{i=1}^{n}\left[g_{i}^{\left[m\right]}-h\left(\bm{\mathrm{x}}_{i};\bm{\xi}^{[m]}\right)\right]^{2},
\end{equation}
where $\left(g_{1}^{\left[m\right]},\cdots,g_{n}^{\left[m\right]}\right)^{\top}$is
the current negative gradient vector of $\Psi$ with respect to (w.r.t.)
$\widehat{F}^{\left[m-1\right]}$:

\begin{equation}
g_{i}^{\left[m\right]}=-\frac{\partial\Psi\left(y_{i},F\left(\bm{\mathrm{x}}_{i}\right),\omega_{i}\right)}{\partial F\left(\bm{\mathrm{x}}_{i}\right)}\bigg|_{F\left(\bm{\mathrm{x}}_{i}\right)=\widehat{F}^{\left[m-1\right]}\left(\bm{\mathrm{x}}_{i}\right)}.
\end{equation}
When fitting this regression trees, first use a fast top-down ``best-fit''
algorithm with a least-squares splitting criterion \citep{friedman2000additive}
to find the splitting variables and the corresponding splitting locations
that determine the terminal regions $\{\widehat{R}_{l}^{[m]}\}_{l=1}^{L}$,
then estimate the terminal-node values $\{\hat{u}_{l}^{[m]}\}_{l=1}^{L}$
. This fitted regression tree $h(\bm{\mathrm{x}};\{\hat{u}_{l}^{[m]},\widehat{R}_{l}^{[m]}\}_{l=1}^{L})$
can be viewed as a tree-constrained approximation of the unconstrained
negative gradient. Due to the disjoint nature of the regions produced
by regression trees, finding the expansion coefficient $\beta^{[m]}$
can be reduced to solving $L$ optimal constants $\eta_{l}^{[m]}$
within each region $\hat{R}_{l}^{[m]}$. And the estimation of $\tilde{F}$
for the next stage becomes
\begin{equation}
\widehat{F}^{[m]}=\widehat{F}^{[m-1]}+\nu\sum_{l=1}^{L}\hat{\eta}_{l}^{[m]}I(\bm{\mathrm{x}}\in\widehat{R}_{l}^{[m]}),\label{TDboost:F_update}
\end{equation}
where $0\leq\nu\leq1$ is the shrinkage factor that controls the update
step size. A small $\nu$ imposes more shrinkage, while $\nu=1$ gives
complete negative gradient steps. \citet{friedman2001greedy} has
found that the shrinkage factor reduces overfitting and improve the
predictive accuracy. The complete algorithm is shown in Algorithm
\ref{algorithm:TDboost}.

\noindent 
\begin{algorithm}
\SetKwData{Left}{left}\SetKwData{This}{this}\SetKwData{Up}{up}
\SetKwFunction{Union}{Union}\SetKwFunction{FindCompress}{FindCompress}
\SetKwInOut{Input}{Input}\SetKwInOut{Output}{Output}

\Input{Dataset $\mathbf{D}=\{(y_{i},\bm{\mathrm{x}}_{i},\omega_{i})\}_{i=1}^{n}$ and the index parameter $\rho$.}
\Output{Estimates $\widehat{F}$.} 

Initialize $\widehat{F}^{[0]}=\log\left(\frac{\sum_{i=1}^{n}\omega_{i}y_{i}}{\sum_{i=1}^{n}\omega_{i}}\right)$.\

\For{$m=0,1,2\ddd M$}{
Compute the negative gradient vector $(g_{1}^{[m]}\ddd g_{n}^{[m]})^{\top}$
$$
g_{i}^{[m]}=\omega_{i}\left\{-y_{i}\exp\left[(1-\rho)\widehat{F}^{[m-1]}(\bm{\mathrm{x}}_{i})\right]+\exp\left[(2-\rho)\widehat{F}^{[m-1]}(\bm{\mathrm{x}}_{i})\right]\right\}, i=1\ddd n.
$$

Fit the negative gradient vector to $(\bm{\mathrm{x}}_{1}\ddd\bm{\mathrm{x}}_{n})^{\top}$ by an $L$-terminal node regression tree, giving the partition $\{\widehat{R}_{l}^{[m]}\}_{l=1}^{L}$.

Compute the optimal terminal node predictions $\eta_{l}^{[m]}$ for each region $\widehat{R}_{l}^{[m]}$, $l=1,2\ddd L$
$$
\hat{\eta}_{l}^{[m]}=\log\left(\frac{\sum_{i:\bm{\mathrm{x}}_{i}\in\widehat{R}_{l}^{[m]}}\omega_{i}y_{i}\exp\left[(1-\rho)\widehat{F}^{[m-1]}(\bm{\mathrm{x}}_{i})\right]}{\sum_{i:\bm{\mathrm{x}}_{i}\in\widehat{R}_{l}^{[m]}}\omega_{i}y_{i}\exp\left[(2-\rho)\widehat{F}^{[m-1]}(\bm{\mathrm{x}}_{i})\right]}\right)
$$

Update $\widehat{F}^{[m]}$ for each region $\widehat{R}_{l}^{[m]}$ by \eqref{TDboost:F_update}.
}
Return $\widehat{F}=\widehat{F}^{[M]}$.

\caption{TDboost Algorithm\label{algorithm:TDboost}}
\end{algorithm}

\subsection{Implementation details\label{subsection:modifed EM}}

\noindent Next we give a data-driven method to find initial values
for parameter estimation. The idea is that we approximately view the
latent variables as $\Pi_{i}\approx I\{y_{i}\neq0\}$. That is, we
treat all zeros as if they are all from the exact zero mass portion,
which can be reasonable for extremely unbalanced zero-inflated data.
If the latent variables were known, it is straightforward to find
the MLE solution of a constant mean model:
\begin{equation}
\boldsymbol{\theta}^{0}=\arg\max_{\boldsymbol{\theta}\in\boldsymbol{\Theta}}\log\mathcal{L}\left(\boldsymbol{\theta};\mathbf{D},\widetilde{\boldsymbol{\Pi}}\right),\label{initial criterion}
\end{equation}
where $\boldsymbol{\Theta}=\mathcal{C}\times\mathbb{R}^{+}\times\left[0,1\right]$
, $\mathcal{C}=\left\{ F\equiv\eta\,|\,\eta\in\mathbb{R}\right\} $,
and $\eta$ is a constant scalar. We then find initial values successively
as follows:

\noindent Initialize $F^{0}$ by
\begin{align}
F^{0}= & \arg\min_{\eta\in\mathbb{R}}\sum_{i=1}^{n}I\{y_{i}\neq0\}\cdot\Psi(y_{i},\eta,\omega_{i})\nonumber \\
= & \log\left[\frac{\sum_{i=1}^{n}I\{y_{i}\neq0\}\cdot y_{i}\cdot\omega_{i}}{\sum_{i=1}^{n}I\{y_{i}\neq0\}\cdot\omega_{i}}\right].\label{initial:F}
\end{align}

\noindent Initialize $\phi^{0}$ by
\begin{equation}
\phi^{0}=\arg\min_{\phi\in\mathbb{R}^{+}}\sum_{i=1}^{n}I\{y_{i}\neq0\}\left(\log a(y_{i},\phi/\omega_{i},\rho)+\frac{\omega_{i}}{\phi}(y_{i}\frac{\exp\left(F^{0}\left(1-\rho\right)\right)}{1-\rho}-\frac{\exp\left(F^{0}\left(2-\rho\right)\right)}{2-\rho})\right),\label{initial:phi}
\end{equation}

\noindent Initialize $q^{0}$ by
\begin{equation}
q^{0}=\frac{1}{n}\sum_{i=1}^{n}I\{y_{i}\neq0\}.\label{initial:q}
\end{equation}

\noindent Given $\boldsymbol{\theta}^{0}$ obtained above, we can
then initialize $(\delta_{1,i},\delta_{0,i})$ by equation \eqref{posterior density1}
and \eqref{posterior density0}, giving $\left(\delta_{0,i}^{0},\delta_{1,i}^{0}\right)$.

As a last note, when implementing EMTboost algorithm, for more stable
computation, we may want to avoid that the probability $q$ converges
to $1$ (or 0). In such case, we can add a regularization term $r\log(1-q)$
on $q$ so that each M-step in Q-function \eqref{Q-function} becomes
\begin{equation}
\mathcal{P}Q\left(\boldsymbol{\theta}|\mathbb{\boldsymbol{\theta}}^{t}\right)=Q\left(\boldsymbol{\theta}|\mathbb{\boldsymbol{\theta}}^{t}\right)+\underbrace{r\log\left(1-q\right)}_{\text{regularization term}},\label{penalized ZIF log-likelihood}
\end{equation}
where $r\in\mathbb{R}^{+}$ is a non-negative regularization parameter.
Apparently, when maximizing the penalized log-likelihood function
\eqref{penalized ZIF log-likelihood}, larger $q$ will be penalized
more. We establish the EM algorithm similar as before, and only need
to modify the Maximization step of \eqref{update:q} w.r.t. $q$:
\begin{equation}
q_{\mathcal{P}}^{t+1}=\frac{\frac{1}{n}\sum_{i=1}^{n}\delta_{1,i}^{t}}{r+1},\label{eq:modified update q}
\end{equation}
pulling the original update $q^{t+1}=\frac{1}{n}\sum_{i=1}^{n}\delta_{1,i}^{t}$
towards $0$ by fraction $r+1$. Alternatively, if in some cases,
we want to avoid that the EMTboost model degrades to an exact zero
mass, the regularization term can be chosen as $r\log\left(1-|1-2q|\right)$.
The updating step with respect to $q$ becomes a soft thresholding
update with the threshold $r$:
\begin{equation}
q_{\mathcal{P}'}^{t+1}=\frac{1}{2}-\frac{\mathcal{S}_{r}\left(1-\frac{2}{n}\sum_{i=1}^{n}\delta_{1,i}^{t}\right)}{2(r+1)}
\end{equation}
where $\mathcal{S}_{r}(\cdot)$ is the soft thresholding function
with $\mathcal{S}_{r}(x)=\text{sign}(x)(|x|-r)_{+}$. We apply these
penalized EMTboost methods to the real data application in Appendix
\ref{Appendix_subsection:real_application_Penalty}.

\section{SIMULATION STUDIES\label{section:Simulation Studies}}

In this section, we compare the EMTboost model (Section \ref{section:EMTboost})
with a regular Tweedie boosting model (that is $q\equiv1$; TDboost)
and the Gradient Tree-Boosted Tobit model (Grabit; \citealp{sigrist2017grabit})
in terms of the function estimation performance. The Grabit model
extends the Tobit model \citep{tobin1958estimation} using gradient-tree
boosting algorithm. We here present two simulation studies in which
zero-inflated data are generated from zero-inflated Tweedie model
(Case 1 in Section \ref{subsection:simulation1}) and zero-inflated
Tobit model (Case 2 in Section \ref{subsection:simulation2}). An
additional simulation result (Case 3) in which data are generated
from a Tweedie model is put in Appendix \ref{Appendix:Simulation3}.

Fitting Grabit, TDboost and EMTboost models to these data sets, we
get the final predictor function $\hat{F}\left(\cdot\right)$ and
parameter estimators. Then we make a prediction about the pure premium
by applying the predictor functions on an independent held-out testing
set to find estimated expectation: $\widehat{\mu}(\textbf{x})=\mathbb{E}\left(y|\textbf{x}\right)$.
For the three competing models, the predicted pure premium is given
by equations 
\begin{align}
\widehat{\mu}^{\text{Grabit}}(\textbf{x}) & =\varphi\left(-\hat{F}_{\text{Grabit}}(\textbf{x})\right)+\hat{F}_{\text{Grabit}}(\textbf{x})\left(1-\Phi\left(-\hat{F}_{\text{Grabit}}(\textbf{x})\right)\right),\label{Grabit expected pure premium}\\
\widehat{\mu}^{\text{TDboost}}(\textbf{x}) & =\exp(\hat{F}_{\text{TDboost}}(\textbf{x})),\\
\widehat{\mu}^{\text{EMTboost}}(\textbf{x}) & =(1-\widehat{q}_{\text{EMTboost}})\exp\left(\hat{F}_{\text{EMTboost}}(\textbf{x})\right),
\end{align}
where $\varphi\left(\cdot\right)$ is the probability density function
of the standard normal distribution and $\Phi(\cdot)$ is its cumulative
distribution function. The predicted pure premium of the Grabit model
is derived in detail in Appendix \ref{Appendix:Tobit}. As the true
model is known in simulation settings, we can campare the difference
between the predicted premiums and the expected true losses. For the
zero-inflated Tobit model in case 2, the expected true loss of this
model is given by $\mathbb{E}_{\text{ZIF-Tobit}}\left[y|F\left(\textbf{x}\right)\right]=q\left[\varphi\left(-F\left(\textbf{x}\right)\right)+F\left(\textbf{x}\right)\left(1-\Phi\left(-F\left(\textbf{x}\right)\right)\right)\right]$,
with $q$ being the probability that response $y$ comes from the
Tobit model and $F\left(\textbf{x}\right)$ the true target funtion.

\subsection{Measurement of Prediction Accuracy\label{subsection:Measurement_of_Prediction_Accuracy}}

Given a portfolio of policies $\mathbf{D}=\{(y_{i},\bm{\mathrm{x_{i}}},\omega_{i})\}_{i=1}^{n}$,
$y_{i}$ is the claim cost for the $i$-th policy and $\hat{y}_{i}$
is denoted as the predicted claim cost. We consider the following
three measurements of prediction accuracy of $\left\{ \hat{y}_{i}\right\} _{i=1}^{n}$.
\begin{description}
\item [{$\textbf{Gini Index (\ensuremath{\text{Gini}^{a}})}\quad$}] Gini
index is a well-accepted tool to evaluate the performance of predictions.
There exists many variants of Gini index and one variant we use is
denoted by $\text{Gini}^{a}$ \citep{ye2018combining}: for a sequence
of numbers $\{s_{1},\cdots,s_{n}\}$, let $R(s_{i})\in\{1,\cdots,n\}$
be the rank of $s_{i}$ in the sequence in an increasing order. To
break the ties when calculating the order, we use the \textbf{LAST}
tie-breaking method, i.e., we set $R(s_{i})>R(s_{j})$ if $s_{i}=s_{j},~i<j$.
Then the normalized Gini index is referred to as:
\begin{equation}
\text{Gini}^{a}=\frac{\frac{\sum_{i=1}^{n}y_{i}R(\hat{y}_{i})}{\sum_{i=1}^{n}y_{i}}-\sum_{i=1}^{n}\frac{n-i+1}{n}}{\frac{\sum_{i=1}^{n}y_{i}R(y_{i})}{\sum_{i=1}^{n}y_{i}}-\sum_{i=1}^{n}\frac{n-i+1}{n}}.
\end{equation}

\noindent Note that this criterion only depends on the rank of the
predictions and larger $\text{Gini}^{a}$ index means better prediction
performance.
\item [{$\textbf{Gini Index (\ensuremath{\text{Gini}^{b}})}\quad$}] We
exploit an popular alternative--the ordered Lorentz curve and the
associated Gini index (denoted by $\text{Gini}^{b}$; \citealp{frees2011summarizing,frees2014insurance})
to capture the discrepancy between the expected premium $P(\textbf{x})=\widehat{\mu}(\textbf{x})$
and the true losses $y$. We successively specify the prediction from
each model as the base premium and use predictions from the remaining
models as the competing premium to compute the $\text{Gini}^{b}$
indices. Let $B(\textbf{x})$ be the ``base premium'' and $P(\textbf{x})$
be the ``competing premium''. In the ordered Lorentz curve, the
distribution of losses and the distribution of premiums are sorted
based on the relative premium $R(\textbf{x})=P(\textbf{x})/B(\textbf{x})$.
The ordered premium distribution is 
\begin{equation}
\hat{D}_{P}(s)=\frac{\sum_{i=1}^{n}B(\textbf{x}_{i})I\left\{ R(\textbf{x}_{i})\leq s\right\} }{\sum_{i=1}^{n}B(\textbf{x}_{i})},
\end{equation}
and the ordered loss distribution is
\begin{equation}
\hat{D}_{L}(s)=\frac{\sum_{i=1}^{n}y_{i}I\left\{ R(\textbf{x}_{i})\leq s\right\} }{\sum_{i=1}^{n}y_{i}}.
\end{equation}
Then the ordered Lorentz curve is the graph of $\left(\hat{D}_{P}(s),\hat{D}_{L}(s)\right)$.
Twice the area between the ordered Lorentz curve and the line of equality
measures the discrepancy between the premium and loss distributions,
and is defined as the $\text{Gini}^{b}$ index.
\item [{$\textbf{Mean Absolute Deviation (MAD)}\quad$}] Mean Absolute
Deviation with respect to the true losses $\left\{ y_{i}\right\} _{i=1}^{n}$
is defined as $\frac{1}{n}\sum_{i=1}^{n}|y_{i}-\hat{y}_{i}|.$ In
the following simulation studies, we can directly compute the mean
absolute deviation between the predicted losses $\left\{ \hat{y}_{i}\right\} _{i=1}^{n}$
and the expected true losses $\left\{ \mathbb{E}\left[y_{i}|\textbf{x}_{i}\right]\right\} _{i=1}^{n}$
to obtain $\frac{1}{n}\sum_{i=1}^{n}|\mathbb{E}\left[y_{i}|\textbf{x}_{i}\right]-\hat{y}_{i}|$,
while in the real data study, we can only compute the MAD against
true losses $\left\{ y_{i}\right\} _{i=1}^{n}$.
\end{description}

\subsection{Case 1\label{subsection:simulation1}}

In this simulation case, we generate data from the zero-inflated Tweedie
models with two different target functions: one with two interactions
and the other generated from \citet{friedman2001greedy}'s ``random
function generator'' (RFG) model. We fit the training data using
Grabit, TDboost, and EMTboost. In all numerical studies, five-fold
cross-validation is adopted to select the optimal ensemble size $M$
and regression tree size $L$, while the shrinkage factor $\nu$ is
set as 0.001.

\subsubsection{Two Interactions Function (Case 1.1)}

In this simulation study, we demonstrate the performance of EMTboost
to recover the mixed data distribution that involves exact zero mass,
and the robustness of our model in terms of premium prediction accuracy
when the index parameter $\rho$ is misspecified. We consider the
true target function with two hills and two valleys: 
\begin{equation}
F(x_{1},x_{2})=e^{-5(1-x_{1})^{2}+x_{2}^{2}}+e^{-5x_{1}^{2}+(1-x_{2})^{2}},\label{two_interactions_target_function}
\end{equation}
which corresponds to a common scenario where the effect of one variable
changes depending on the effect of the other. The response $Y$ follows
a zero-inflated Tweedie distribution $\text{ZIF-Tw}(\mu,\phi,\rho,q)$
with Tweedie portion probability $q$:
\begin{align}
Y\sim\left\{ \begin{array}{ll}
Z, & \text{with probability }q,\ \text{Z\ensuremath{\sim}Tw}(\mu,\phi,\rho),\\
0, & \text{with probability }1-q,
\end{array}\right.\label{ZIF-Tw distribution}
\end{align}
where 
\begin{align*}
\mu=\exp(F(x_{1},x_{2}))~,x_{1},x_{2}\text{\ensuremath{\overset{\text{ind}.}{\sim}}}\text{Unif}(0,1),
\end{align*}
with $\phi=1$, $\rho=1.5$ and $q$ chosen from a decreasing sequence
of values: $q\in\{1$, $0.85$, $0.75$, $0.50$, $0.25$, $0.10\}$.

We generate $n=500$ observations $\{\textbf{x}_{i},y_{i}\}_{i=1}^{n}$
for training and $n'=1200$ for testing, and fit the training data
using Grabit, TDboost and EMTboost models. The true target functions
are known, and we use MAD (against expected true premium) and $\text{Gini}^{a}$
index as performance criteria.

When fitting EMTboost, we design three scenarios to illurstrate the
robustness of our method w.r.t. $\rho$. In the first scenario, set
$\rho=1.5$, which is the true value. In the second scenario, set
$\rho=1.7$, which is misspecified. In the last scenario, we use the
profile likelihood method to estimate $\rho$. 

\begin{figure}[H]
\includegraphics[scale=0.48]{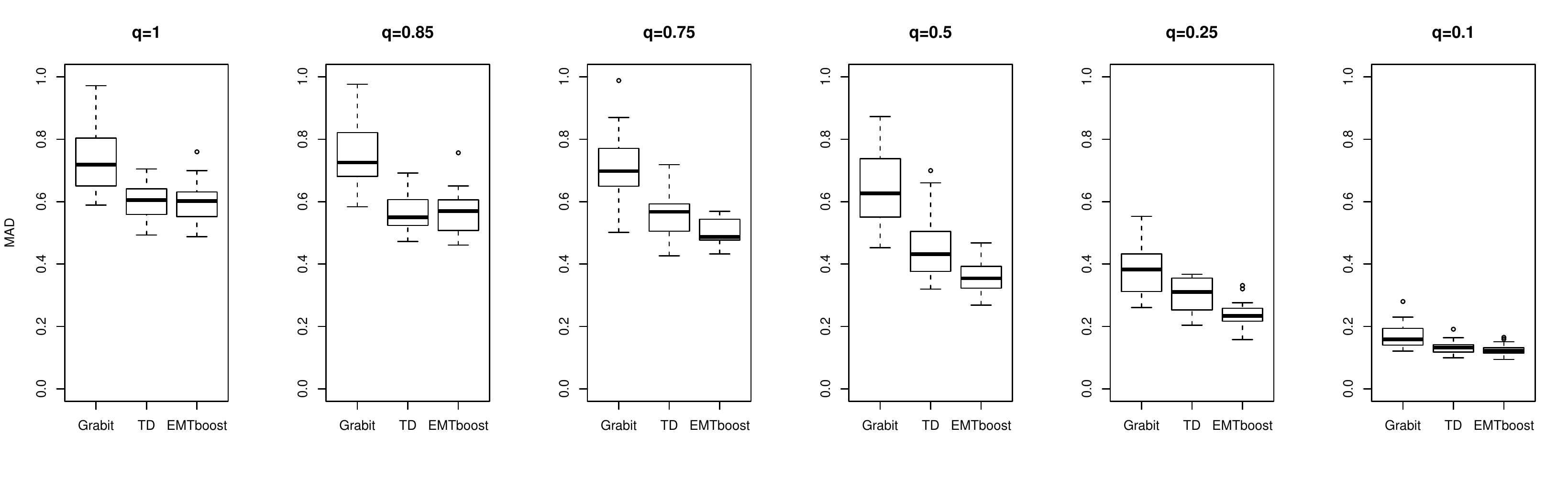}

\noindent \caption{Simulation results for case 1.1: comparing MADs of Grabit, TDboost
and EMTboost with decreasing $q$. Boxplots display empirical distributions
of MADs based on $20$ independent replications. \label{fig:Sim1.1_Grabit_TD_EMTboost_MAD} }
\end{figure}

The resulting MADs and $\text{Gini}^{a}$ indices of the three competing
models on the held-out testing data are reported in Table \ref{table:Sim1_TD_Grabit_EMTboost_MAD}
and Table \ref{table:Sim1_TD_Grabit_EMTboost_GINI^a}, which are averaged
over $20$ independent replications for each $q$. Boxplots of MADs
comparing Grabit, TDboost and EMTboost (with estimated $\rho$) are
shown in Figure \ref{fig:Sim1.1_Grabit_TD_EMTboost_MAD}. In all three
scenarios, EMTboost outperforms Grabit and TDboost in terms of the
ability to recover the expected true premium by giving smallest MADs
and largest $\text{Gini}^{a}$ indices, especially when zeros inflate:
$q\in\left\{ 0.5,0.25,0.1\right\} $. The prediction performance of
EMTboost when $\rho=1.7$ is not much worse than that when $\rho=1.5$,
showing that the choice of $\rho$ has relatively small effect on
estimation accuracy. 

\noindent 
\begin{table}[H]
\centering{}\caption{Simulation results for case 1.1 with MADs. \label{table:Sim1_TD_Grabit_EMTboost_MAD} }
\begin{tabular}{cccccc}
\toprule 
 & \multicolumn{5}{c}{Competing Models}\tabularnewline
\cmidrule{2-6} 
\multirow{2}{*}{$q$} & \multirow{2}{*}{TDboost} & \multirow{2}{*}{Grabit} & \multicolumn{3}{c}{EMTboost}\tabularnewline
 &  &  & $\rho=1.5$ & $\rho=1.7$ & tuned $\rho$\tabularnewline
$1.00$ & $0.597\,(.013)$ & $0.746\,(.029)$ & $0.594\,(.016)$ & $0.598\,(.012)$ & $0.598\,(.015)$\tabularnewline
$0.85$ & $0.565\,(.015)$ & $0.761\,(.032)$ & $0.554\,(.017)$ & $0.555\,(.017)$ & $0.562\,(.016)$\tabularnewline
$0.75$ & $0.561\,(.018)$ & $0.706\,(.026)$ & $0.489\,(.010)$ & $0.485\,(.011)$ & $0.503\,(.010)$\tabularnewline
$0.50$ & $0.454\,(.024)$ & $0.674\,(.044)$ & $0.365\,(.012)$ & $0.375\,(.014)$ & $0.361\,(.012)$\tabularnewline
$0.25$ & $0.301\,(.013)$ & $0.382\,(.019)$ & $0.240\,(.010)$ & $0.242\,(.011)$ & $0.237\,(.010)$\tabularnewline
$0.10$ & $0.135\,(.005)$ & $0.169\,(.009)$ & $0.122\,(.004)$ & $0.124\,(.004)$ & $0.124\,(.004)$\tabularnewline
\bottomrule
\end{tabular}
\end{table}

\noindent 

\noindent 
\begin{table}[H]
\centering{}\begin{center}
\caption{Simulation results for case 1.1 with $\text{Gini}^{a}$ indices. \label{table:Sim1_TD_Grabit_EMTboost_GINI^a} }
\begin{tabular}{cccccc}
\toprule 
 & \multicolumn{5}{c}{Competing Models}\tabularnewline
\cmidrule{2-6} 
\multirow{2}{*}{$q$} & \multirow{2}{*}{TDboost} & \multirow{2}{*}{Grabit} & \multicolumn{3}{c}{EMTboost}\tabularnewline
 &  &  & $\rho=1.5$ & $\rho=1.7$ & tuned $\rho$\tabularnewline
$1.00$ & $0.480\,(.008)$ & $0.449\,(.011)$ & \textbf{$0.481\,(.006)$} & \textbf{$0.481\,(.006)$} & \textbf{$0.481\,(.006)$}\tabularnewline
$0.85$ & $0.393\,(.008)$ & $0.354\,(.009)$ & \textbf{$0.397\,(.007)$} & \textbf{$0.397\,(.007)$} & \textbf{$0.397\,(.007)$}\tabularnewline
$0.75$ & $0.343\,(.009)$ & $0.300\,(.020)$ & \textbf{$0.363\,(.008)$} & \textbf{$0.365\,(.007)$} & \textbf{$0.361\,(.008)$}\tabularnewline
$0.50$ & $0.242\,(.012)$ & $0.186\,(.016)$ & \textbf{$0.289\,(.011)$} & \textbf{$0.288\,(.012)$} & \textbf{$0.292\,(.011)$}\tabularnewline
$0.25$ & $0.172\,(.016)$ & $0.116\,(.020)$ & \textbf{$0.219\,(.016)$} & \textbf{$0.215\,(.017)$} & \textbf{$0.217\,(.015)$}\tabularnewline
$0.10$ & $0.085\,(.028)$ & $0.107\,(.023)$ & \textbf{$0.137\,(.027)$} & \textbf{$0.122\,(.028)$} & \textbf{$0.136\,(.025)$}\tabularnewline
\bottomrule
\end{tabular}
\par\end{center}
\end{table}

\subsubsection{Random Function Generator (Case 1.2) \label{subsubsection:sim1_Random-Function-Generator}}

In this case, we compare the performance of the three competing models
in various complicated and randomly generated predictor functions.
We use the RFG model whose true target function $F$ is randomly generated
as a linear expansion of functions $\left\{ g_{k}\right\} _{k=1}^{20}$
:
\begin{equation}
F\left(\textbf{x}\right)=\sum_{k=1}^{20}b_{k}g_{k}\left(\textbf{z}_{k}\right).
\end{equation}
Here, each coefficient $b_{k}$ is a uniform random variable from
$\text{Unif\ensuremath{\left[-1,1\right]}}$. Each $g_{k}\left(\textbf{z}_{k}\right)$
is a function of $\textbf{z}_{k}$, where $\textbf{z}_{k}$ is defined
as a $p_{k}$-sized subset of the $p$-dimensional variable $\textbf{x}$
in the form
\begin{equation}
\textbf{z}_{k}=\left\{ x_{\psi_{k}\left(j\right)}\right\} _{j=1}^{p_{k}}.
\end{equation}
where each $\psi_{k}$ is an independent permutation of the integers
$\left\{ 1,\cdots,p\right\} $. The size $p_{k}$ is randomly selected
by $\min\left(\lfloor2.5+r_{k}\rfloor,p\right)$, where $r_{k}$ is
generated from an exponential distribution with mean $2$. Hence,
the expected order of interaction presented in each $g_{k}\left(\textbf{z}_{k}\right)$
is between four and five. Each function $g_{k}\left(\textbf{z}_{k}\right)$
is a $p_{k}$-dimensional Gaussian function:
\begin{equation}
g_{k}\left(\textbf{z}_{k}\right)=\exp\left\{ -\frac{1}{2}\left(\textbf{z}_{k}-\textbf{u}_{k}\right)^{\text{T}}\textbf{V}_{k}\left(\textbf{z}_{k}-\textbf{u}_{k}\right)\right\} ,
\end{equation}
where each mean vector $\textbf{u}_{k}$ is randomly generated from
$\text{N}\left(0,\textbf{I}_{p_{k}}\right)$. The $p_{k}\times p_{k}$
covariance matrix $\textbf{V}_{k}$ is defined by 
\begin{equation}
\textbf{V}_{k}=\textbf{U}_{k}\textbf{D}_{k}\textbf{U}_{k}^{\text{T}},
\end{equation}
where $\textbf{U}_{k}$ is a random orthonormal matrix, $\textbf{D}_{k}=\text{diag}\left\{ d_{k}\left[1\right],\cdots,d_{k}\left[p_{k}\right]\right\} $,
and the square root of each diagonal element $\sqrt{d_{k}\left[j\right]}$
is a uniform random variable from $\text{Unif}\left[0.1,2.0\right]$.
We generate data $\left\{ y_{i},\textbf{x}_{i}\right\} _{i=1}^{n}$
from zero-inflated Tweedie distribution where $\textbf{x}_{i}\sim\text{N}\left(0,\textbf{I}_{p}\right)$,
$\mu_{i}=\exp\left\{ F\left(\textbf{x}_{i}\right)\right\} $, $i=1,\cdots,n$.

\begin{figure}[H]
\includegraphics[scale=0.48]{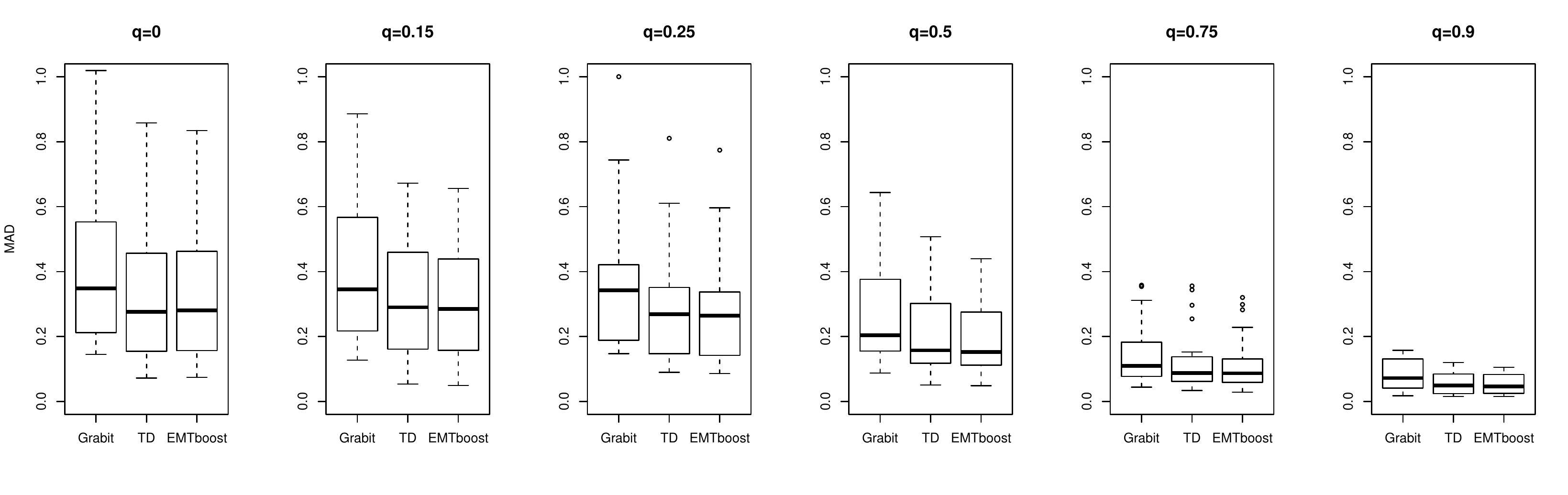}

\noindent \caption{Simulation results for case 1.2: comparing MADs of Grabit, TDboost
and EMTboost with decreasing $q$. Boxplots display empirical distributions
of the MADs based on $20$ independent replications. \label{fig:Sim1.2_Grabit_TD_EMTboost_MAD} }
\end{figure}

We randomly generate $20$ sets of samples with $\phi=1$ and $\rho=1.5$,
each sample having $2000$ observations, $1000$ for training and
$1000$ for testing. When fitting EMTboost for each $q\in\left\{ 1,0.85,0.75,0.5,0.25,0.1\right\} $,
the estimates of Tweedie portion probability have mean $\bar{q^{*}}=0.96,0.79,0.71,0.53,0.28,0.13$.
Figure \ref{fig:Sim1.2_Grabit_TD_EMTboost_MAD} shows simulation results
comparing the MADs of Grabit, TDboost and EMTboost. We can see, in
all the cases, EMTboost outperforms Grabit and becomes very competitive
compared to TDboost when $q$ decreases. 

\noindent 

\subsection{Case 2\label{subsection:simulation2}}

In this simulation case, we generate data from the zero-inflated Tobit
models with two target functions similar to that of case 1. For all
three gradient-tree boosting models, five-fold cross-validation is
adopted for developing trees. Profile likelihood method is used again.

\subsubsection{Two Interactions Function (Case 2.1)\label{subsubsection:sim2.1}}

In this simulation study, we compare the performance of three models
in terms of MADs. Consider the data generated from the zero-inflated
Tobit model where the true target function is given by 
\begin{align}
F(x_{1},x_{2})=2\cos\left(2.4\pi(|x_{1}|^{3}+|x_{2}|^{3})^{0.5}\right).\label{target_func:Tobit_simulation}
\end{align}
Conditional on covariates $\textbf{X}=(X_{1},X_{2})$, the latent
variable $Y^{*}$ follows a Gaussian distribution:
\begin{align*}
Y^{*}=F(X_{1},X_{2})+\epsilon,~X_{k}\text{ i.i.d.}\sim\text{Unif}(-1,1),~k=1,2,~\epsilon\sim N(0,1).
\end{align*}
The Tobit response $Y_{\text{Tobit}}$ can be expressed as $Y_{\text{Tobit}}=\max(Y^{*},0)$,
and we generate the zero-inflated Tobit data using the Tobit response:
\begin{equation}
Y\sim\begin{cases}
Y_{\text{Tobit}}, & \text{with probability }q,\\
0, & \text{with probability }1-q.
\end{cases}\label{ZIF-Tobit model}
\end{equation}
where $q$ takes value from the sequence $\left\{ 1,0.85,0.75,0.5,0.25,0.1\right\} $.

\begin{figure}[H]
\begin{centering}
\includegraphics[scale=0.48]{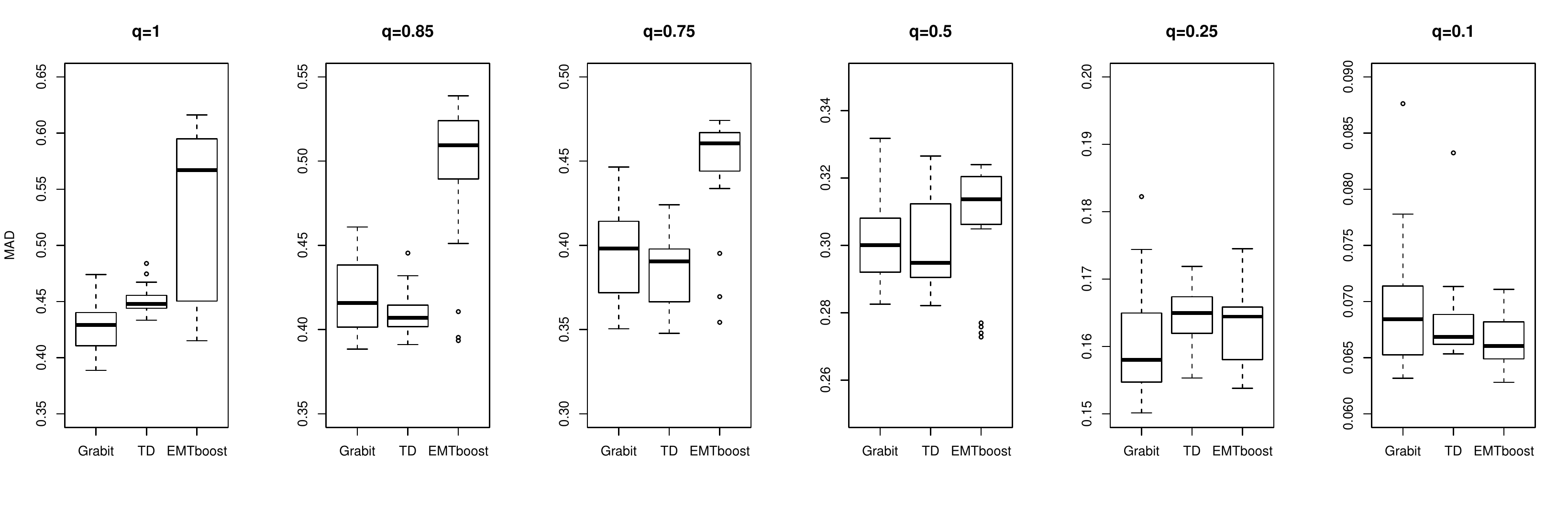}
\par\end{centering}
\centering{}\caption{Simulation results for case 2.1: comparing the MADs of Grabit, TDboost
and EMTboost with decreasing $q$. Boxplots display empirical distributions
of MADs based on $20$ independent replications. \label{fig:Sim2.1_Grabit_TD_EMTboost_MAD} }
\end{figure}

We generate $n=500$ observations for training and $n'=4500$ for
testing. Figure \ref{fig:Sim2.1_Grabit_TD_EMTboost_MAD} shows simulation
results when comparing MADs of Grabit, TDboost and EMTboost based
on $20$ independent replications. We can see from the first boxplot
that when $q=1$, zero-inflated Tobit distribution degenerates to
a Tobit distribution, and not surprisingly, Grabit outperforms EMTboost
in MADs. As $q$ decreases, meaning the proportion of zeros increases,
the prediction performance of EMTboost gets improved. When the exact
zero mass probability is $1-q=0.9$, the averaged MADs of the three
models are $\overline{\text{MAD}}_{\text{Grabit}}=0.0697$, $\overline{\text{MAD}}_{\text{TDboost}}=0.0681$,
$\overline{\text{MAD}}_{\text{EMTboost}}=0.0664$, with EMTboost performing
the best. 

\noindent 

\subsubsection{Random Function Generator (Case 2.2)}

\begin{figure}[H]
\begin{centering}
\includegraphics[scale=0.46]{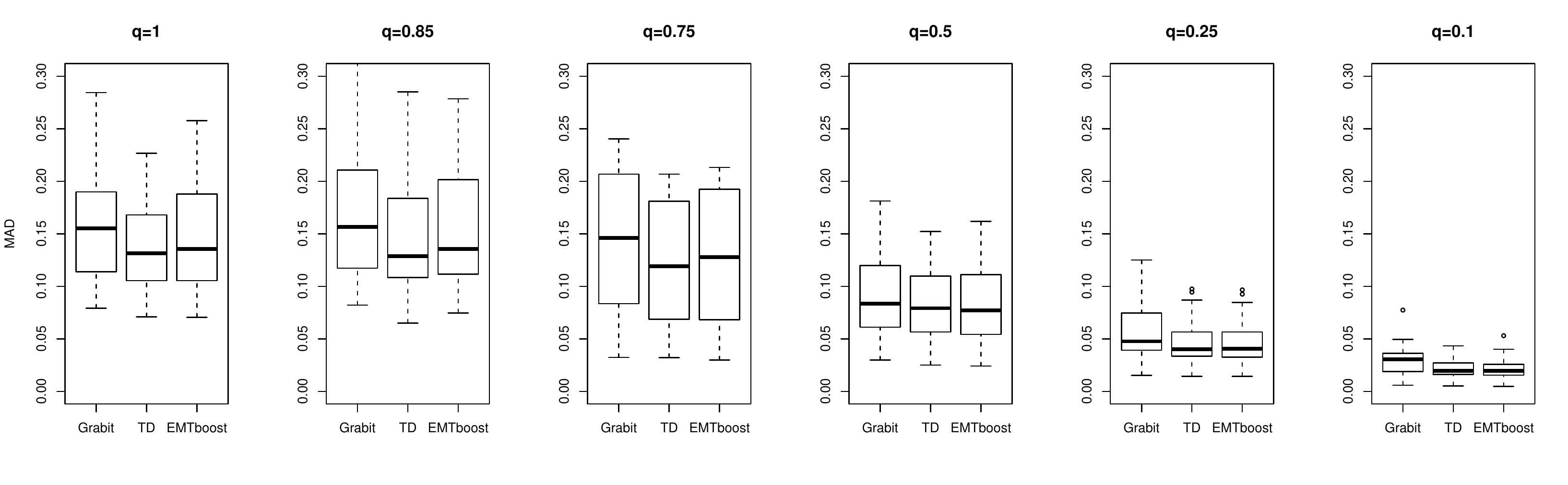}
\par\end{centering}
\centering{}\caption{Simulation results for case 2.2: comparing the MADs of Grabit, TDboost
and EMTboost when decreasing $q$. Boxplots display empirical distributions
of MADs based on $20$ independent replications. \label{fig:Sim2.2_Grabit_TD_EMTboost_MAD} }
\end{figure}

We again use the RFG model in this simulaiton. The true target function
$F$ is randomly generated as given in Section \ref{subsubsection:sim1_Random-Function-Generator}.
The latent variable $Y^{*}$ follows
\begin{align*}
Y^{*}=F(\textbf{x}_{i})+\epsilon,~\textbf{x}_{i}\sim\text{N}\left(0,\textbf{I}_{p}\right),~\epsilon\sim\text{N}(0,1),\;i=1,\cdots,n.
\end{align*}
We set $Y_{\text{Tobit}}=\max(Y^{*},0)$ and generate the data following
the zero-inflated Tobit model \eqref{ZIF-Tobit model} with Tobit
portion probability $q\in\left\{ 1,0.85,0.75,0.5,0.25,0.1\right\} $.
We randomly generate $20$ sets of sample from the zero-inflated Tobit
model for each $q$, and each sample contains $2000$ observations,
$1000$ for training and $1000$ for testing. Figure \ref{fig:Sim2.2_Grabit_TD_EMTboost_MAD}
shows MADs of Grabit, TDboost and EMTboost as boxplots. Interestingly,
for all the $q$'s, TDboost and EMTboost outperform Grabit even though
the true model is a Tobit model. The MAD of EMTboost becomes better
when $q$ decreases, and is competitive with that of TDboost when
$q=0.1$: the averaged MADs of the three models are $\overline{\text{MAD}}_{\text{Grabit}}=0.0825$,
$\overline{\text{MAD}}_{\text{TDboost}}=0.0565$, $\overline{\text{MAD}}_{\text{EMTboost}}=0.0564$.
As for the averaged $\text{Gini}^{a}$ indices, EMTboost performs
the best when $q=0.1$: $\overline{\text{Gini}^{a}}_{\text{Grabit}}=0.0463$,
$\overline{\text{Gini}^{a}}_{\text{TDboost}}=0.0816$, $\overline{\text{Gini}^{a}}_{\text{EMTboost}}=0.1070$.

\noindent 

\section{APPLICATION: AUTOMOBILE CLAIMS\label{section:application:real data}}

\subsection{Data Set}

We consider the auto-insurance claim data set as analyzed in \citet{yip2005modeling}
and \citet{zhang2005boosting}. The data set contains 10,296 driver
vehicle records, each including an individual driver's total claim
amount ($z_{i}$) and 17 characteristics $\textbf{x}_{i}=(x_{i,1},\cdots,x_{i,17})$
for the driver and insured vehicle. We want to predict the expected
pure premium based on $\textbf{x}_{i}$. The description statistics
of the data are provided in \citet{yang2017insurance}. Approximately
$61.1\%$ of policyholders had no claims, and $29.6\%$ of the policyholders
had a positive claim amount up to $\$10,000$. Only $9.3\%$ of the
policy-holders had a high claim amount above $\$10,000$, but the
sum of their claim amount made up to $64\%$ of the overall sum. We
use this original data set to synthesize the often more realistic
scenarios with extremely unbalanced zero-inflated data sets. 

Specifically, we randomly under-sample (without replacement) from
the nonzero-claim data with certain fraction $\lambda$ to increase
the percentage of the zero-claim data. For example, if we set the
under-sampling fraction as $\lambda=0.15$, then the percentage of
the non-claim policyholders will become approximately $61.1/(61.1+38.9\lambda)=91.28\%$.
We choose a decreasing sequence of under-sampling fractions $\lambda\in\{1,0.75,0.5,0.25,0.15,0.1\}$.
For each $\lambda$, we randomly under-sample the positive-loss data
without replacement and combine these nonzero-loss data with the zero-loss
data to generate a new data set. Then we separate this new data set
into two sets uniformly for training and testing . The corresponding
percentages of zero-loss data among the new data set w.r.t. different
$\lambda$ are presented in Table \ref{table:real_Zero percentage}.
The Grabit, TDboost and EMTboost models are fitted on the training
set and their estimators are obtained with five-fold cross-validation. 

\begin{table}[H]
\centering{}\caption{Real - Zero percentage w.r.t. $\lambda$.\label{table:real_Zero percentage}}
\begin{tabular}{cccccccc}
\toprule 
$\lambda$ & $1$ & $0.75$ & $0.50$ & $0.25$ & $0.15$ & $0.10$ & $0.05$\tabularnewline
\cmidrule{2-8} 
Zero Percentage & $61.1\%$ & $67.7\%$ & $75.9\%$ & $86.3\%$ & $91.3\%$ & $94.0\%$ & $96.9\%$\tabularnewline
\bottomrule
\end{tabular} 
\end{table}

\subsection{Performance Comparison}

To compare the performance of Grabit, TDboost and EMTboost models,
we predict the pure premium $P(\textbf{x})$ by applying each model
on the held-out testing set. Since the losses are highly right-skewed,
we use the orderd Lorentz curve and the associated $\text{Gini}^{b}$
index described in Section \ref{subsection:Measurement_of_Prediction_Accuracy}
to capture the discrepancy between the expected premiums and true
losses. 

\begin{figure}[H]
\begin{centering}
\includegraphics[scale=0.53]{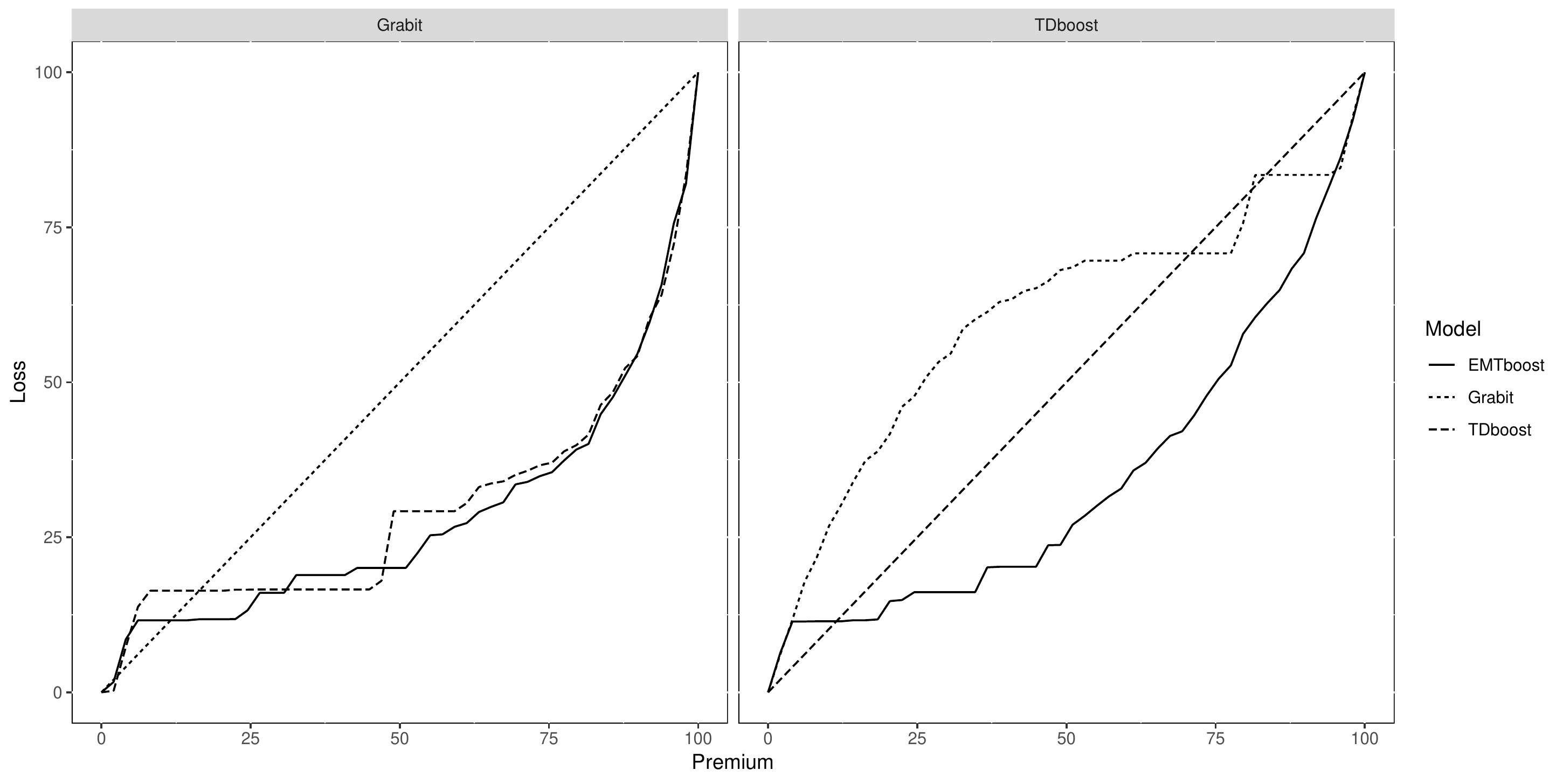}
\par\end{centering}
\centering{}\caption{The ordered Lorentz curves for the synthetic data on a single replication
when $\lambda=0.15$. Grabit (or TDboost) is set as the base premium
and the EMTboost is the competing premium. The ordered Lorentz curve
of EMTboost is below the line of equality when choosing Grabit or
TDboost as the base premium.\label{fig:real_LorentzCurves} }
\end{figure}

The entire procedure of under-sampling, data separating and $\text{Gini}^{b}$
index computation are repeated 20 times for each $\lambda$. A sequence
of matrices of the averaged $\text{Gini}^{b}$ indices and standard
errors w.r.t. each under-sampling fraction $\lambda$ are presented
in Table \ref{table:Real(total)_Grabit_TD_EMTboost_GINI^b}. We then
follow the ``minimax'' strategy \citep{frees2014insurance} to pick
the ``best'' base premium model that is least vulnerable to the
competing premium models. For example, when $\lambda=0.15$, the maximal
$\text{Gini}^{b}$ index is $40.381$ when using $B(\textbf{x})=\widehat{\mu}^{\text{Grabit}}(\textbf{x})$
as the base premium, $36.735$ when $B(\textbf{x})=\widehat{\mu}^{\text{TDboost}}(\textbf{x})$
, and $-22.674$ when $B(\textbf{x})=\widehat{\mu}^{\text{EMTboost}}(\textbf{x})$.
Therefore, EMTboost has the smallest maximum $\text{Gini}^{b}$ index
at $-22.674$, hence having the best performance. Figure \ref{fig:real_LorentzCurves}
also shows that when Grabit (or TDboost) is selected as the base premium,
EMTboost represents the most favorable choice.

\noindent 

\begin{figure}[H]
\begin{centering}
\includegraphics[scale=0.6]{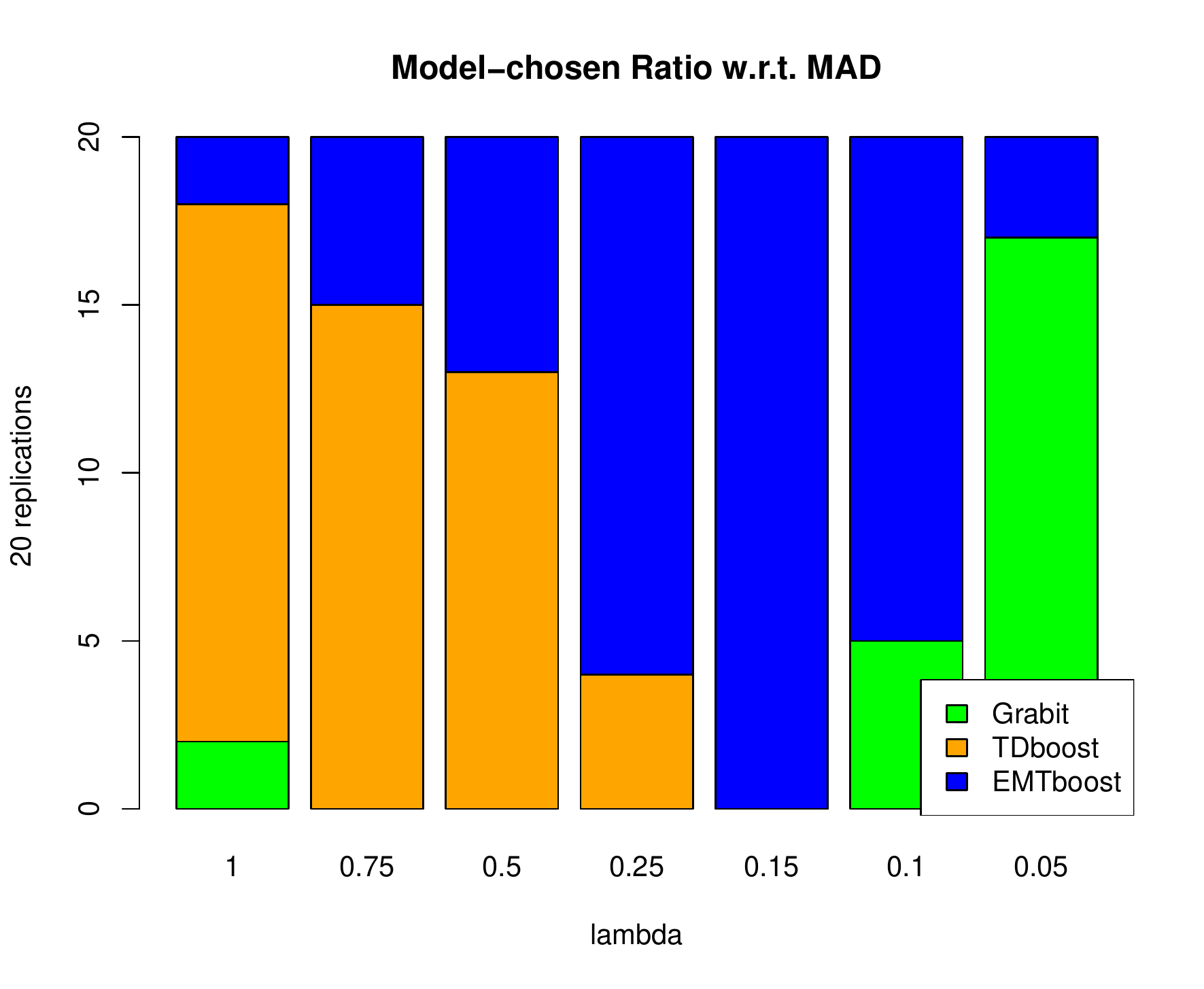}
\par\end{centering}
\centering{}\caption{Barplots of model-selection ratio among Grabit, TDboost and EMTboost
w.r.t. $\text{Gini}^{b}$ indices under $20$ independent replications
when $\lambda$ increases. \label{fig:real_model-chosen RATIO} }
\end{figure}

After computing the $\text{Gini}^{b}$ index matrix and using the
``minimax'' strategy to choose the best candidate model, we count
the frequency, out of 20 replications, of each model chosen as the
best model and record the ratio of their frequencies. The results
w.r.t each $\lambda$ are demenstrated in Figure \ref{fig:real_model-chosen RATIO}.
From Table \ref{table:Real(total)_Grabit_TD_EMTboost_GINI^b} and
Figure \ref{fig:real_model-chosen RATIO}, we find that when $\lambda$
decreases, the performance of EMTboost gradually outperforms that
of TDboost in terms of averaged $\text{Gini}^{b}$ indices and the
corresponding model-selection ratios. In particular, TDboost outperforms
EMTboost when $\lambda=1,0.75,0.5$, and EMTboost outperforms TDboost
when $\lambda=0.25,0.15,0.1,0.05.$ When $\lambda=0.25,0.15,0.1$,
EMTboost has the largest model-selection ratio among the three. 

\begin{table}[H]
\centering{}\caption{Grabit, TDboost and EMTboost $\text{Gini}^{b}$ indices. The ``best''
base premium models are emphasized based on the matrices of averaged
$\text{Gini}^{b}$ indices. \label{table:Real(total)_Grabit_TD_EMTboost_GINI^b}}
\begin{tabular}{rrrr}
\toprule 
 & \multicolumn{3}{c}{Competing Premium}\tabularnewline
\cmidrule{2-4} 
Base Premium & Grabit & TDboost & EMTboost\tabularnewline
\midrule
 & \multicolumn{3}{c}{$\lambda=1$}\tabularnewline
\cmidrule{2-4} 
Grabit & $0$ & $11.459\,(0.417)$ & $11.402\,(0.386)$\tabularnewline
$\textbf{TDboost}$ & $\bm{6.638}\,(0.409)$ & $0$ & $0.377\,(0.355)$\tabularnewline
EMTboost & $7.103\,(0.357)$ & $2.773\,(0.414)$ & $0$\tabularnewline
\midrule 
 & \multicolumn{3}{c}{$\lambda=0.75$}\tabularnewline
\cmidrule{2-4} 
Grabit & $0$ & $14.955\,(0.413)$ & $15.162\,(0.435)$\tabularnewline
$\textbf{TDboost}$ & $\bm{5.466}\,(0.425)$ & $0$ & $1.848\,(0.504)$\tabularnewline
EMTboost & $6.152\,(0.385)$ & $2.622\,(0.560)$ & $0$\tabularnewline
\midrule 
 & \multicolumn{3}{c}{$\lambda=0.50$}\tabularnewline
\cmidrule{2-4} 
Grabit & $0$ & $25.047\,(1.539)$ & $25.621\,(1.492)$\tabularnewline
\textbf{$\textbf{TDboost}$} & $3.516\,(0.963)$ & $0$ & \textbf{$\bm{4.056}\,(0.651)$}\tabularnewline
EMTboost & $5.702\,(0.698)$ & $2.525\,(0.501)$ & $0$\tabularnewline
\midrule 
 & \multicolumn{3}{c}{$\lambda=0.25$}\tabularnewline
\cmidrule{2-4} 
Grabit & $0$ & $51.502\,(1.062)$ & $51.581\,(1.005)$\tabularnewline
TDboost & $-18.248\,(2.445)$ & $0$ & $20.035\,(2.414)$\tabularnewline
$\textbf{EMTboost}$ & $1.283\,(2.593)$ & $\bm{3.929}\,(2.544)$ & $0$\tabularnewline
\midrule 
 & \multicolumn{3}{c}{$\lambda=0.15$}\tabularnewline
\cmidrule{2-4} 
Grabit & $0$ & $37.290\,(2.505)$ & $40.381\,(1.822)$\tabularnewline
TDboost & $-23.569\,(2.607)$ & $0$ & $36.735\,(3.188)$\tabularnewline
$\textbf{EMTboost}$ & $\bm{-22.674}\,(1.975)$ & $-22.926\,(2.604)$ & $0$\tabularnewline
\midrule 
 & \multicolumn{3}{c}{$\lambda=0.10$}\tabularnewline
\cmidrule{2-4} 
Grabit & $0$ & $-1.189\,(5.828)$ & $16.721\,(5.120)$\tabularnewline
TDboost & $14.581\,(6.587)$ & $0$ & $35.298\,(3.026)$\tabularnewline
$\textbf{EMTboost}$ & \textbf{$\bm{-2.742}\,(4.884)$} & $-20.080\,(3.572)$ & $0$\tabularnewline
\midrule 
 & \multicolumn{3}{c}{$\lambda=0.05$}\tabularnewline
\cmidrule{2-4} 
$\textbf{Grabit}$ & $0$ & $-16.851\,(2.662)$ & \textbf{$\bm{-8.652}\,(3.059)$}\tabularnewline
TDboost & $42.493\,(3.792)$ & $0$ & $27.754\,(3.784)$\tabularnewline
EMTboost & $32.169\,(3.767)$ & $-13.448\,(3.551)$ & $0$\tabularnewline
\bottomrule
\end{tabular} 
\end{table}

\noindent 
\begin{table}[H]
\begin{centering}
\caption{Comparing Grabit, TDboost and EMTboost with MADs.\label{table:real_Grabit_TD_EMTboost_MAD}}
\par\end{centering}
\centering{}%
\begin{tabular}{cccc}
\toprule 
 & \multicolumn{3}{c}{Competing Models}\tabularnewline
\cmidrule{2-4} 
$\lambda$ & Grabit & TDboost & EMTboost\tabularnewline
$1.00$ & $4.248\,(.014)$ & $4.129\,(.012)$ & $4.067\,(.012)$\tabularnewline
$0.75$ & $3.879\,(.017)$ & $3.679\,(.011)$ & $3.622\,(.012)$\tabularnewline
$0.50$ & $3.345\,(.026)$ & $2.994\,(.017)$ & $2.928\,(.016)$\tabularnewline
$0.25$ & $2.439\,(.014)$ & $1.945\,(.021)$ & \textbf{$1.766\,(.014)$}\tabularnewline
$0.15$ & $1.720\,(.015)$ & $1.489\,(.023)$ & \textbf{$1.309\,(.019)$}\tabularnewline
$0.10$ & $1.265\,(.011)$ & $1.100\,(.015)$ & $0.986\,(.014)$\tabularnewline
$0.05$ & $0.714\,(.012)$ & $0.578\,(.015)$ & $0.402\,(.009)$\tabularnewline
\bottomrule
\end{tabular}
\end{table}

We also find that TDboost and EMTboost both outperform Grabit when
$\lambda=1$, $0.75$, $0.5$, $0.25$, $0.15$, $0.1$, but Grabit
becomes the best when $\lambda=0.05$; interestingly, if we compare
MAD results in Table \ref{table:real_Grabit_TD_EMTboost_MAD}, the
prediction error of EMTboost becomes the smallest for each $\lambda$.
This inconsistent results between the criteria MAD and $\text{Gini}^{b}$
index when $\lambda=0.05$ can be explained by the different learning
characteristics of the EMTboost methods and the Grabit methods. To
see it more clearly, we compute the MADs on the positive-loss dataset,
denoted by $\text{MAD}^{+}$, and zero-loss dataset, denoted by $\text{MAD}^{0}$
seperately, and compute the $\text{Gini}^{a}$ indices on the nonzero
dataset, denoted by $\text{Gini}^{a+}$. When $\lambda=0.05$, EMTboost
obtains the smallest averaged MAD on zero-loss dataset ($\overline{\text{MAD}^{0}}_{\text{EMTboost}}^{\text{}}=0.146$
\textless{} $\overline{\text{MAD}^{0}}_{\text{TDboost}}=0.269$ \textless{}
$\overline{\text{MAD}^{0}}_{\text{Grabit}}=0.422$), while Grabit
obtains the smallest averaged MAD on positive-loss dataset ($\overline{\text{MAD}^{+}}_{\text{EMTboost}}=10.406$
\textgreater{} $\overline{\text{MAD}^{+}}_{\text{TDboost}}=10.297$
\textgreater{} $\overline{\text{MAD}^{+}}_{\text{Grabit}}=9.927$).
The $\text{MAD}^{0}$ performance shows that EMTboost captures the
zero information much better than TDboost and Grabit. The somewhat
worse $\text{MAD}^{+}$ performance of EMTboost when $\lambda=0.05$
can be explained by the deficiency of the nonzero data points (only
about $100$ nonzeros comparing with over $3000$ zeros); if we fix
the nonzero sample size with under-sampling fraction $\lambda=0.2$,
and at the same time, over-sample the zero-loss part with over-sampling
fraction $\eta=3$ to obtain about $96\%$ zero proportion, then the
averaged $\text{Gini}^{b}$ results summarized in Table \ref{table:real_Grabit_TD_EMTboost_under_0.2_over_3_penal1_GINI^b}
in Appendix \ref{Appendix_subsection:real_under_0.2_over_3} indeed
show that EMTboost remains to perform competitively compared with
the other methods under this large zero proportion setting. 

\section{Concluding Remarks\label{section:Conclusions}}

We have proposed and studied the EMTboost model to handle very unbalanced
claim data with excessive proportion of zeros. Our proposal overcomes
the difficulties that traditional Tweedie model have when handling
these common data scenarios, and at the same time, preserves the flexibility
of nonparametric models to accomdate complicated nonlinear and high-order
interaction relations. We also expect that our zero-inflated Tweedie
approach can be naturally extended to high-dimensional linear settings
\citep{qian2016tweedie}. It remains interesting to develop extended
approaches to subject-specific zero-inflation settings, and provide
formal procedures that can conveniently test if zero-inflated Tweedie
model is necessary in data analysis compared to its simplified alternatives
under both parametric and nonparametric frameworks.

\bibliographystyle{asa}

\appendix

\section{Appendix A: Tobit Model (Truncated Normal Distribution)\label{Appendix:Tobit}}

Suppose the latent variable $Y^{*}$ follows, conditional on covariate
$\textbf{x}$, a Gaussian distribution:
\begin{equation}
Y^{*}|\textbf{x}\sim N(\mu(\textbf{x}),\sigma^{2})
\end{equation}

\noindent This latent variable $Y^{*}$ is observed only when it lies
in an interval $\left[y_{l},y_{u}\right]$. Otherwise, one observes
$y_{l}$ or $y_{u}$ depending on whether the latent variable is below
the lower threshold $y_{l}$ or above the upper threshold $y_{u}$,
respectively. Denoting $Y$ as the observed variable, we can express
it as:
\begin{equation}
Y=\begin{cases}
y_{l}, & \text{if }Y^{*}\leq y_{l},\\
Y^{*}, & \text{if }y_{l}<Y^{*}<y_{u},\\
y_{u}, & \text{if }y_{u}\leq Y^{*}.
\end{cases}
\end{equation}

\noindent The density of $Y$ is given by:
\begin{align}
f_{\text{Tobit}}\left(y;\mu(\textbf{x}),\sigma\right) & =\Phi\left(\frac{y_{l}-\mu(\textbf{x})}{\sigma}\right)\textbf{I}_{y_{l}}(y)+\left(1-\Phi\left(\frac{y_{u}-\mu(\textbf{x})}{\sigma}\right)\right)\textbf{I}_{y_{u}}(y).\nonumber \\
 & +\frac{1}{\sigma}\varphi\left(\frac{y-\mu(\textbf{x})}{\sigma}\right)\textbf{I}\left\{ y_{l}<y<y_{u}\right\} 
\end{align}

\noindent Then the expectation of $Y|\textbf{x}$ is given by: 
\begin{align}
\mathbb{E}_{\sigma}\left[y|\textbf{x}\right] & =\int_{-\infty}^{+\infty}yf_{\text{Tobit}}\left(y;\mu(\textbf{x}),\sigma\right)dy\nonumber \\
 & =y_{l}\Phi\left(\alpha\right)+\int_{y_{l}}^{y_{u}}y\frac{1}{\sigma}\varphi\left(\frac{y-\mu(\textbf{x})}{\sigma}\right)dy+y_{u}\left(1-\Phi\left(\beta\right)\right)\nonumber \\
 & =y_{l}\Phi\left(\alpha\right)+\int_{\alpha}^{\beta}(s\sigma+\mu(\textbf{x}))\varphi\left(s\right)ds+y_{u}\left(1-\Phi\left(\beta\right)\right)\nonumber \\
 & =y_{l}\Phi\left(\alpha\right)+\sigma\left(\varphi\left(\alpha\right)-\varphi\left(\beta\right)\right)+\mu(\textbf{x})\left(\Phi\left(\beta\right)-\Phi\left(\alpha\right)\right)+y_{u}\left(1-\Phi\left(\beta\right)\right),\\
\nonumber 
\end{align}

\noindent where $\alpha=\frac{y_{l}-\mu(\textbf{x})}{\sigma},\beta=\frac{y_{u}-\mu(\textbf{x})}{\sigma}$.
And
\begin{equation}
\phi(\xi)=\frac{1}{\sqrt{2\pi}}\exp\left(-\frac{1}{2}\xi^{2}\right)
\end{equation}

\noindent is the probability density function of the standard normal
distribution and $\Phi(\cdot)$ is its cumulative distribution function:
\begin{equation}
\Phi(y)=\int_{-\infty}^{y}\varphi(\xi)d\xi
\end{equation}

In simulation 2, the latent variable is truncated by $0$ from below,
i.e., $y_{l}=0,y_{u}=\infty$. So we have $\varphi(\beta)=0,\Phi(\beta)=1$.
We also set the variance of the Gaussian distribution as $\sigma=1$.
Then the expectation of $Y|\textbf{x}$ is given by:
\begin{equation}
\mathbb{E}_{\sigma=1}\left[y|\text{\textbf{x}}\right]=\varphi\left(-\mu(\textbf{x})\right)+\mu(\textbf{x})\left(1-\Phi\left(-\mu(\textbf{x})\right)\right).
\end{equation}

\section{Appendix B: Case 3\label{Appendix:Simulation3}}

In this simulation study, we demonstrate that our EMTboost model can
fit the nonclaim dataset well. We consider the data generated from
the Tweedie model, with the true target function \eqref{two_interactions_target_function}.
We generate response $Y$ from the Tweedie distribution $\text{Tw}(\mu,\phi,\rho)$,
with $\mu=\exp(F(x_{1},x_{2}))~,x_{1},x_{2}\sim\text{Unif}(0,1)$
and the index parameter $\rho=1.5$. We find that when the dispersion
parameter $\phi$ takes large value in $\mathbb{\mathbb{R}^{+}}$,
Tweedie's zero mass probability $\mathbb{P}(Y_{\text{Tw}}=0)=\exp(-\frac{1}{\phi}\frac{\mu^{2-\rho}}{2-\rho})$
gets closer to $1$. So we choose three large dispersion values $\phi\in\left\{ 20,30,50\right\} $.

We generate $n=500$ observations $\{\textbf{x}_{i},y_{i}\}_{i=1}^{n}$
for training and $n'=1000$ for testing, and fit the training data
using Grabit, TDboost and EMTboost models. For all three models, five-fold
cross-validation is adopted and the shrinkage factor $\nu$ is set
as $0.001$. The profile likelihood method is used. 

The discrepancy between the predicted loss and the expected true loss
in criteria MAD and $\text{Gini}^{b}$ index are shown in Table \ref{table:Sim3_Grabit_TD_EMTboost_MAD}
and Table \ref{table:Sim3_Grabit_TD_EMTboost_GINI^b}, which are averaged
over 20 independent replications for each $\phi$. Table \ref{table:Sim3_Grabit_TD_EMTboost_MAD}
shows that EMTboost obtains the smallest MAD. In terms of $\text{Gini}^{b}$
indices, EMTboost is also chosen as the ``best'' model for each
$\phi$. 

In this setting, $\mathbb{P}(Y_{\text{Tw}}=0)\approx0.83,0.88,0.91$,
which means that all the customers are generally very likely to have
no claim. The assumption of our EMTboost model with a general exact
zero mass probability coincides with this data structure. Its zero
mass probability estimation $1-\widehat{q}$ is $0.863$, $0.909$
and $0.943$ respectively, showing that EMTboost learns this zero
part of information quite well. As a result, EMTboost performs no
worse than TDboost, which is based on the true model assumption. 

\noindent 
\begin{table}[H]
\centering{}\caption{Simulation results for case 3 with MADs.\label{table:Sim3_Grabit_TD_EMTboost_MAD} }
\begin{tabular}{cccc}
\toprule 
 & \multicolumn{3}{c}{Competing Models}\tabularnewline
\cmidrule{2-4} 
$\phi$ & Grabit & TDboost & EMTboost\tabularnewline
$20$ & $2.499\,(.072)$ & $2.465\,(.057)$ & \textbf{$2.449\,(.054)$}\tabularnewline
$30$ & $2.442\,(.068)$ & $2.492\,(.060)$ & \textbf{$2.442\,(.060)$}\tabularnewline
$50$ & $2.553\,(.091)$ & $2.560\,(.082)$ & \textbf{$2.544\,(.080)$}\tabularnewline
\bottomrule
\end{tabular}
\end{table}

\noindent 
\begin{table}[H]
\centering{}\caption{Simulation results for case 3 with $\text{Gini}^{b}$ indices. \label{table:Sim3_Grabit_TD_EMTboost_GINI^b} }
\begin{tabular}{rrrr}
\toprule 
 & \multicolumn{3}{c}{Competing premium}\tabularnewline
\cmidrule{2-4} 
Base premium & Grabit & TDboost & EMTboost\tabularnewline
\midrule
 & \multicolumn{3}{c}{$\phi=20$}\tabularnewline
\cmidrule{2-4} 
Grabit & $0$ & $8.205\,(3.492)$ & $8.052\,(3.471)$\tabularnewline
TDboost & $4.487\,(2.462)$ & $0$ & $4.104\,(2.283)$\tabularnewline
\textbf{EMTboost} & \textbf{$\bm{3.153}\,(1.982)$} & $2.347\,(1.656)$ & $0$\tabularnewline
\midrule
 & \multicolumn{3}{c}{$\phi=30$}\tabularnewline
\cmidrule{2-4} 
Grabit & $0$ & $4.273\,(3.447)$ & $3.244\,(3.411)$\tabularnewline
TDboost & $4.017\,(3.257)$ & $0$ & $3.454\,(3.191)$\tabularnewline
\textbf{EMTboost} & \textbf{$\bm{1.404}\,(2.730)$} & $0.846\,(2.403)$ & $0$\tabularnewline
\midrule
 & \multicolumn{3}{c}{$\phi=50$}\tabularnewline
\cmidrule{2-4} 
Grabit & $0$ & $5.452\,(4.806)$ & $10.362\,(4.801)$\tabularnewline
TDboost & $3.479\,(3.654)$ & $0$ & $2.476\,(3.899)$\tabularnewline
\textbf{EMTboost} & $-0.836\,(2.785)$ & \textbf{$\bm{1.031}\,(3.354)$} & $0$\tabularnewline
\bottomrule
\end{tabular}
\end{table}

\section{Appendix C: Real Data\label{Appendix:real}}

\subsection{Implemented EMTboost: Penalization on $q$\label{Appendix_subsection:real_application_Penalty}}

When fitting the EMTboost model, we want to avoid the situation that
the Tweedie portion probability estimation $\widehat{q}$ degenerates
to $0$. So we add a regularization term $r\log\left(1-q\right)$
to the Q-function, as equation \eqref{eq:modified update q} in Section
\ref{subsection:modifed EM}. We choose an increasing sequence of
penalty parameter $\log_{10}\left(r\right)\in\left\{ -2,\cdots,1\right\} _{41}$,
and use the strategy of ``warm start'' to improve the computation
efficiency, i.e., setting the current solution $\left(\widehat{\mu}\left(r_{l}\right),\widehat{\phi}\left(r_{l}\right),\widehat{q}\left(r_{l}\right)\right)$
as the initialization for the next solution $\left(\widehat{\mu}\left(r_{l+1}\right),\widehat{\phi}\left(r_{l+1}\right),\widehat{q}\left(r_{l+1}\right)\right)$. 

\begin{figure}[H]
\begin{centering}
\includegraphics[scale=0.55]{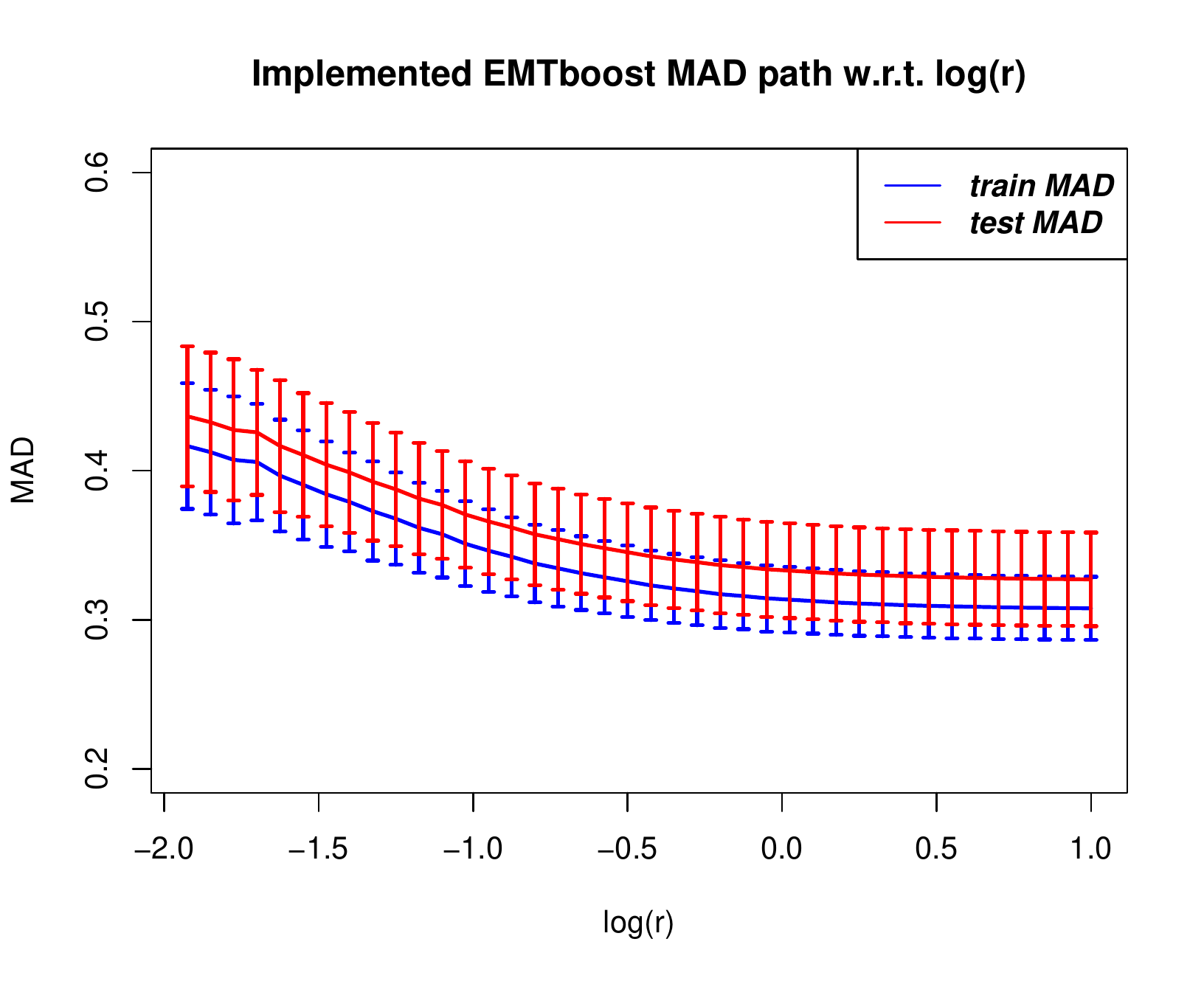}
\par\end{centering}
\begin{centering}
\includegraphics[scale=0.6]{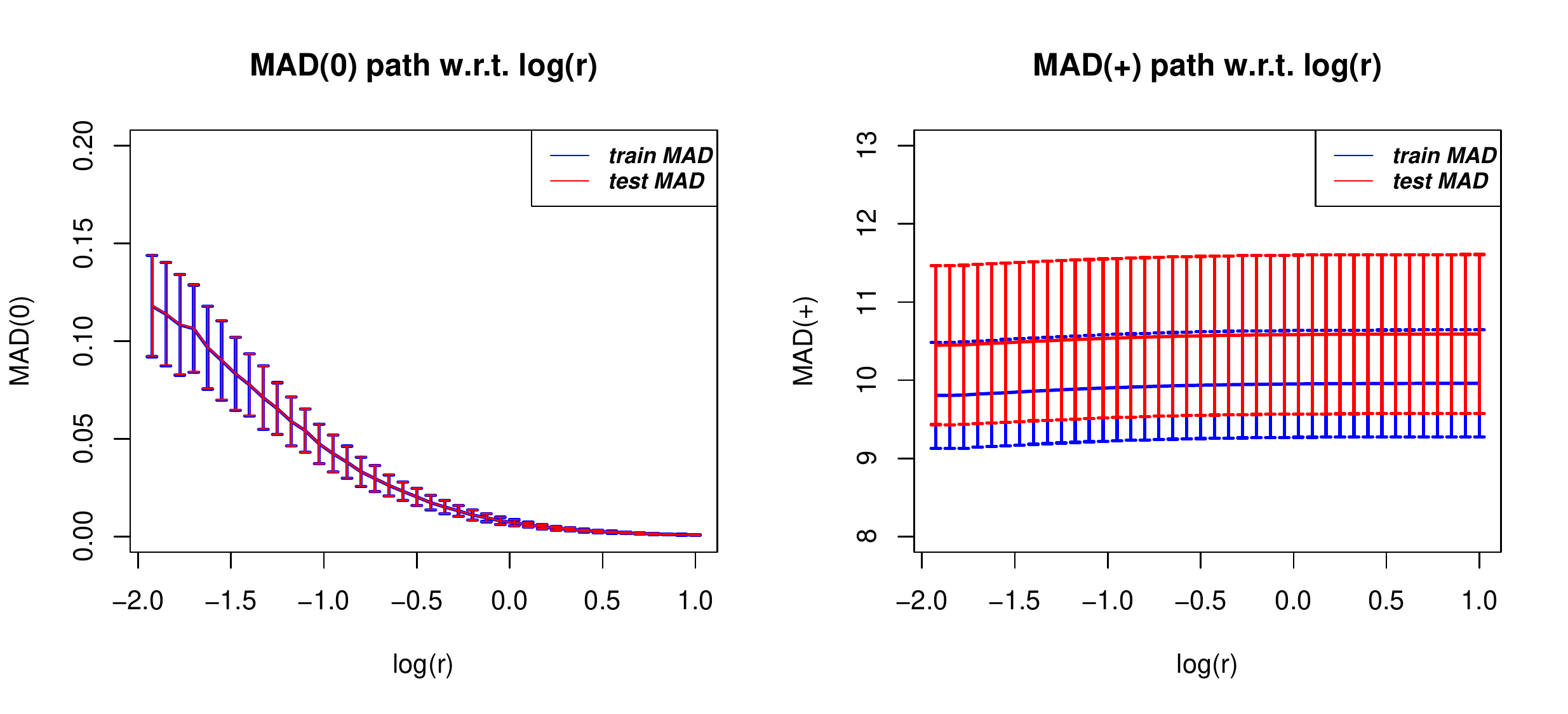}
\par\end{centering}
\centering{}\caption{Implemented EMTboost ($\lambda=0.05$) results: MAD error path with
its one standard deviation when increasing regularization parameter
$r$. The blue lines are the training error lines, and the red ones
are the testing error lines. Top figure: averaged MAD drops when penalty
parameter increases. Bottom left: averaged $\text{MAD}^{0}$ and its
standard deviation drop remarkably and approximate $0$ when $r$
increases. Bottom right: averaged $\text{MAD}^{+}$ is flat and increases
a little when $r$ increases. All the averaged testing MADs are within
one standard deviation of the averaged training MADs. \label{fig:real_EMTboost_penal1_MAD_path} }
\end{figure}

We use this implemented EMTboost model to train the penalized solution
paths with respect to the sequence of penalty parameter $r$ on the
extremely zero-inflated training data $\left(\lambda=0.05\right)$
under $20$ independent replications, and then apply the estimators
to the testing data to compute the discrepancy under MAD. We also
compute the MADs on zero-loss dataset ($\text{MAD}^{0}$) and positive-loss
dataset ($\text{MAD}^{+}$) seperately, and the $\text{Gini}^{a}$
indices on positive-loss dataset ($\text{Gini}^{a+}$). Table \ref{table:real_EMTboost_penal1_path_MAD}
and Figure \ref{fig:real_EMTboost_penal1_MAD_path} shows the implemented
EMTboost MAD path w.r.t. the logrithm of a sequence of penalty parameter
$r$. 

\begin{table}[H]
\begin{centering}
\caption{Implemented EMTboost ($\lambda=0.05$) results with $\text{MAD}$s,
$\text{MAD}^{0}$s and $\text{MAD}^{+}$s. \label{table:real_EMTboost_penal1_path_MAD} }
\par\end{centering}
\centering{}%
\begin{tabular}{lrrrrr}
\toprule 
$\log(r)$ & $-2.00$  & $-1.10$  & $-0.65$  & $0.25$  & $0.70$ \tabularnewline
\cmidrule{2-6} 
$\text{MAD}$ & $0.432\,(0.047)$ & \textbf{$0.377\,(0.036)$} & \textbf{$0.351\,(0.033)$} & $0.330\,(0.032)$ & $0.328\,(0.031)$\tabularnewline
$\text{MAD}^{0}$ & $0.114\,(0.026)$ & $0.054\,(0.011)$ & \textbf{$0.026\,(0.005)$} & $0.004\,(0.001)$ & $0.002\,(0.000)$\tabularnewline
$\text{MAD}^{+}$ & \textbf{$10.452\,(1.017)$} & $10.527\,(1.017)$ & $10.560\,(1.017)$ & $10.587\,(1.017)$ & $10.590\,(1.017)$\tabularnewline
\bottomrule
\end{tabular}
\end{table}

\noindent 

Table \ref{table:real_Grabit_TD_EMTboost_penal1_zero_nonzero_result}
shows the results of Grabit, TDboost, EMTboost, and implemented EMTboost
$\left(r=1/6\right)$. Implemented EMTboost performs the worst in
$\text{Gini}^{a}$, $\text{Gini}^{a+}$ and $\text{MAD}^{+}$, but
the best in MAD and $\text{MAD}^{0}$.

\begin{table}[H]
\begin{centering}
\caption{Comparing Grabit, TDboost, EMTboost and Implemented EMTboost ($r=1/6$)
with $\text{Gini}^{a}$ indices and MADs. \label{table:real_Grabit_TD_EMTboost_penal1_zero_nonzero_result}}
\par\end{centering}
\centering{}%
\begin{tabular}{lrrrr}
\toprule 
$\lambda=0.05$ & Grabit & TDboost & EMTboost & Implemented EMTboost\tabularnewline
\midrule 
$\widehat{q}$ & - & - & $0.697\,(.018)$ & $0.132\,(.003)$\tabularnewline
$\text{Gini}^{a}$ & $0.415\,(.023)$ & $0.134\,(.040)$ & $0.238\,(.033)$ & $0.261\,(.033)$\tabularnewline
$\text{Gini}^{a+}$ & \textbf{$0.104\,(.030)$} & $-0.116\,(.039)$ & $-0.164\,(.031)$ & $-0.103\,(.040)$\tabularnewline
$\text{MAD}$ & $0.714\,(.012)$ & $0.578\,(.015)$ & \textbf{$0.482\,(.013)$} & \textbf{$0.356\,(.008)$}\tabularnewline
$\text{MAD}^{0}$ & $0.422\,(.009)$ & $0.269\,(.012)$ & $0.146\,(.008)$ & \textbf{$0.032\,(.002)$}\tabularnewline
$\text{MAD}^{+}$ & \textbf{$9.927\,(.223)$} & $10.287\,(.226)$ & $10.406\,(.228)$ & $10.551\,(.227)$\tabularnewline
\bottomrule
\end{tabular}
\end{table}

\subsection{Resampleing: Under-sampling Fraction $\lambda=0.2$ and Over-sampling
Fraction $\eta=3$\label{Appendix_subsection:real_under_0.2_over_3}}

We control the nonzero sample size in real application by under-sampling
the nonzero-loss data set with fraction $\lambda=0.2$ and over-sampling
the zero-loss data with fraction $\eta=3$, generating a data set
containing $95.9\%$ zeros. Following the training and testing procedure
in Section \ref{section:application:real data}, Table \ref{table:real_Grabit_TD_EMTboost_under_0.2_over_3_penal1_GINI^b}shows
that EMTboost has the smallest of the maximal (averaged) $\text{Gini}^{b}$
indices, thus is chosen as the ``best'' model. 

\begin{table}[H]
\centering{}\caption{Grabit, TDboost, EMTboost $\text{Gini}^{b}$ indices with $\lambda=0.2$
and $\eta=3$. \label{table:real_Grabit_TD_EMTboost_under_0.2_over_3_penal1_GINI^b}}
\begin{tabular}{rrrr}
\toprule 
$\lambda=0.2,\eta=3$ & \multicolumn{3}{c}{Competing Premium}\tabularnewline
\cmidrule{2-4} 
Base Premium & Grabit & TDboost & EMTboost\tabularnewline
\midrule
Grabit & $0$ & $23.872\,(3.928)$ & $35.616\,(2.655)$\tabularnewline
TDboost & $-4.822\,(2.822)$ & $0$ & $27.234\,(2.152)$\tabularnewline
\textbf{EMTboost} & \textbf{$\bm{-5.583}\,(1.599)$} & $-14.806\,(1.681)$ & $0$\tabularnewline
\bottomrule
\end{tabular} 
\end{table}

\end{document}